\documentclass[journal]{IEEEtran}
\usepackage{cite}

\usepackage{bbm}
\usepackage{color,soul}
\usepackage{gensymb}
\usepackage{dsfont}

\ifCLASSINFOpdf
  \usepackage[pdftex]{graphicx}
  \graphicspath{{../pdf/}{../jpeg/}}
  \DeclareGraphicsExtensions{.pdf,.jpeg,.png}
\else
  \usepackage[dvips]{graphicx}
  \graphicspath{{../eps/}}
  \DeclareGraphicsExtensions{.eps}
\fi
\usepackage[cmex10]{amsmath}
\usepackage{amssymb}
\DeclareMathOperator*{\argmaxA}{arg\,max}

\usepackage{euscript}

\usepackage{multirow}
\usepackage{algorithm}
\usepackage[]{algpseudocode}
\usepackage{diagbox}
\usepackage{physics}
\usepackage{tikz}
\definecolor{myred}{RGB}{205 38 38}

\ifCLASSOPTIONcompsoc
  \usepackage[caption=false,font=normalsize,labelfont=sf,textfont=sf]{subfig}
\else
  \usepackage[caption=false,font=footnotesize]{subfig}
\fi
\usepackage{url}

\begin{document}
%
\title{Learning Latent Interactions for Event classification via Graph Neural Networks and PMU Data}
%
%
%

\author{Yuxuan~Yuan,~\IEEEmembership{Graduate Student Member,~IEEE,}
    Zhaoyu~Wang,~\IEEEmembership{Senior Member,~IEEE,}
	and Yanchao~Wang
\thanks{This work is supported by the U.S. Department of Energy Office of Electricity under DEOE0000910 (\textit{Corresponding author: Zhaoyu Wang})

Y. Yuan, Z. Wang, and Y. Wang are with the Department of Electrical and Computer Engineering, Iowa State University, Ames,
IA 50011 USA (e-mail: yuanyx@iastate.edu; wzy@iastate.edu).
 }
}
%
%

\markboth{}%
{Shell \MakeLowercase{\textit{et al.}}: Bare Demo of IEEEtran.cls for Journals}
%



\maketitle

\begin{abstract}
Phasor measurement units (PMUs) are being widely installed on power systems, providing a unique opportunity to enhance wide-area situational awareness. One essential application is the use of PMU data for real-time event identification. However, how to take full advantage of all PMU data in event identification is still an open problem. Thus, we propose a novel method that performs event identification by mining interaction graphs among different PMUs. The proposed interaction graph inference method follows an entirely data-driven manner without knowing the physical topology. Moreover, unlike previous works that treat interactive learning and event identification as two different stages, our method learns interactions jointly with the identification task, thereby improving the accuracy of graph learning and ensuring seamless integration between the two stages. Moreover, to capture multi-scale event patterns, a dilated inception-based method is investigated to perform feature extraction of PMU data. To test the proposed data-driven approach, a large real-world dataset from tens of PMU sources and the corresponding event logs have been utilized in this work. Numerical results validate that our method has higher classification accuracy compared to previous methods.

\end{abstract}

\begin{IEEEkeywords}
Event identification, interaction graph inference, phasor measurement units, graph neural network.
\end{IEEEkeywords}

\section{Introduction}\label{introduction}
Power systems are in need of better situational awareness due to the integration of new technologies such as distributed renewable generation and electric vehicles. Recently, a rapid growth in the number of phasor measurement units (PMUs) has been observed in power systems. In the U.S., by the end of 2017, the number of recorded PMUs was about 1,900, which is a nine-fold growth from 2009. Compared to the traditional power system monitoring devices, PMUs provide highly granular (e.g., 30 or 60 samples per second) and synchronized measurements, including voltage and current phasor, frequency, and frequency variation, which enables capturing most dynamics of power systems. Hence, researchers and practitioners are exploring a variety of methods to use PMU data for enhancing system monitoring and control. One of the important applications is real-time event identification, which is directly related to event analysis \cite{JD2010}. Event identification models trained on PMU data not only provide supervisory monitoring, but also maintain partial system awareness when supervisory control and data acquisition (SCADA) is dysfunctional, as was the case during the 2003 North American large-scale blackout \cite{SB2017}.

In recent years, a number of papers have explored data-driven methods for event identification to enhance situational awareness of power systems using PMU data. The previous works in this area can be broadly classified into two categories based on the number of PMUs used for model development: \textit{Class I}: each PMU is treated independently, and a single PMU data stream for each event is assigned as one data sample \cite{MC2019,TX2015,DK2017,mk2018,XL2014,EP2008,YG2015}. In \cite{MC2019}, a signal processing-based methodology consisting of the swinging door trending algorithm and dynamic programming was proposed to identify power events. In \cite{TX2015}, principal component analysis (PCA) was used to detect abnormal system behavior and adopt system visualizations. In \cite{DK2017}, by using PMU data in Korea, a wavelet-based event classification model was developed by observing the difference between voltage and frequency signals. In \cite{mk2018}, an empirical model decomposition was utilized to assess power system events using wide-area post-event records. In \cite{XL2014}, an online event detection algorithm was developed based on the change of core subspaces of the PMU data at the occurrence of an event. In \cite{EP2008}, the extended Kalman-filtering algorithm was applied to detect and classify voltage events. In \cite{YG2015}, a knowledge-based criterion was proposed to classify power system events. \textit{Class II}: Instead of using data from a single PMU, several papers perform event classification tasks using multiple PMU measurements, which integrate interactive relationships of different PMUs \cite{MB2016,SL2020,WL2018,RM2018,NSS2017,ZYX2019}. In these methods, the data of each event that includes multiple PMU data streams is assigned as one data sample for model development. In \cite{MB2016}, a scheme was proposed for supervisory protection and situational awareness, which presented a new metric to identify PMU with the strongest signature and an extreme learning machine-based event classifier. In \cite{SL2020}, a data-driven algorithm was proposed, which consists of an unequal-interval method for dimensionality reduction and a PCA-based search method for event detection. The basic idea is to measure similarities and local outlier factors between any two PMU data streams. In \cite{WL2018}, a data-driven event classification method was proposed by characterizing an event utilizing a low-dimensional row subspace spanned by the dominant singular vectors of a high-dimensional spatial-temporal PMU data matrix. In \cite{RM2018}, a correlation-based method was developed to concurrently monitor multiple PMU data streams for identifying system events. In \cite{NSS2017}, an event characterization algorithm was presented using computation of spectral kurtosis on sum of intrinsic mode functions. In \cite{ZYX2019}, a new nonparametric learning framework was proposed for the novelty detection problem with multiple correlated time series by extending the classical smoothness and fitness optimization. A summary of the literature is shown in Table \ref{table:form}.

\begin{table*}
\begin{center}
 \caption{Available Literature on Data-driven Event Classification in Power Systems}
  \includegraphics[width=1\linewidth]{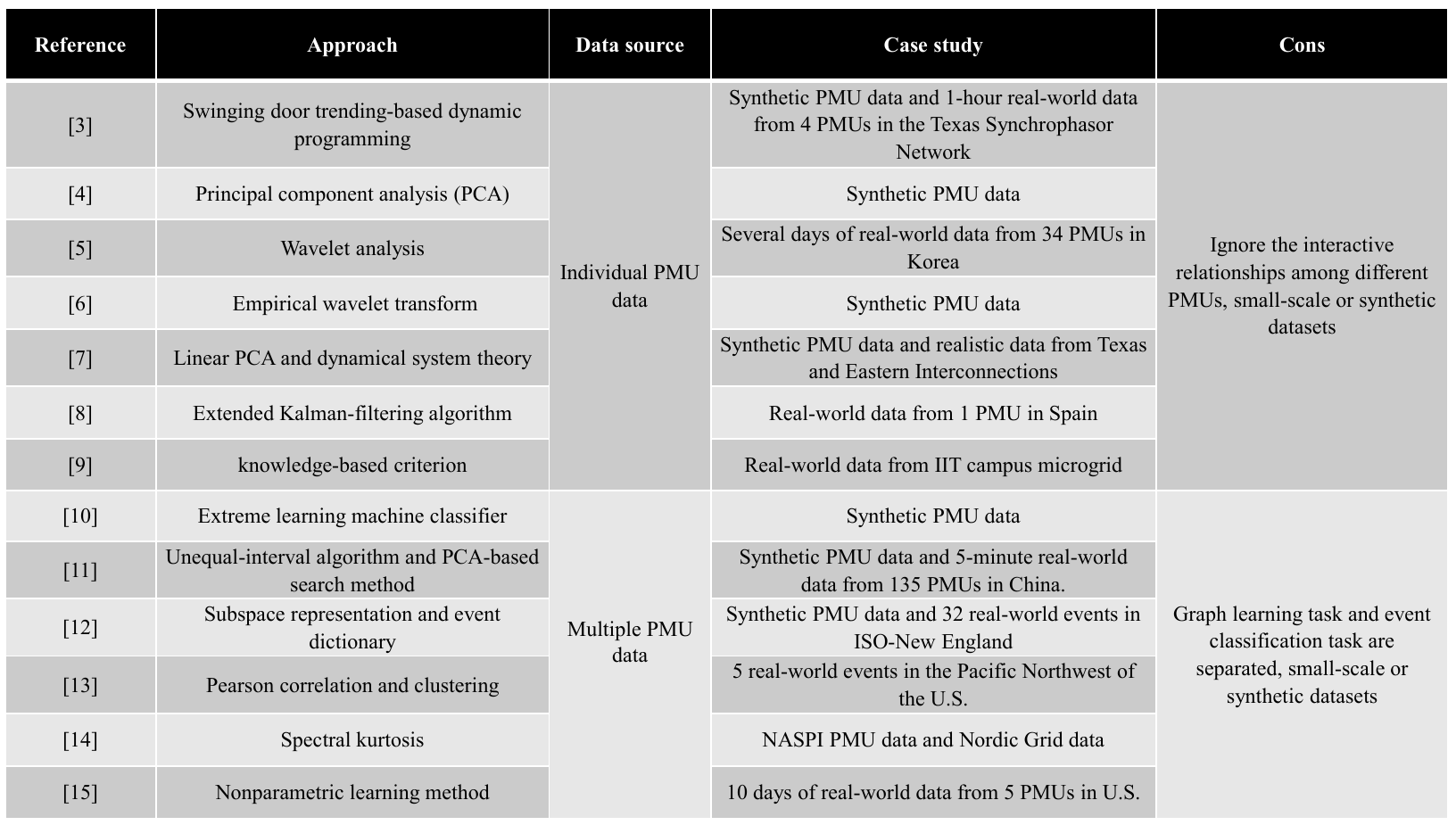}
    \label{table:form}
\end{center}
 \vspace{-2em}
\end{table*}


While these methods have led to meaningful guidelines and invaluable insights, some questions remain open with respect to real-time PMU-based event identification. Basically, Class I models focus on analyzing events using data recorded by individual PMUs. This indicates that the interactive relationship among different PMUs are simply ignored. When applying these event identification models to the actual grids, some PMUs may report events whereas others report normal, resulting in conflicting opinions due to data heterogeneity. On the other hand, Class II methods are generally based on a simplifying assumption that each PMU has the same interactive relationship with the rest of PMUs. This means representing the interaction with a fully connected graph. However, such an assumption may not be realistic due to the complexity of power systems. A natural solution to this problem is to apply statistical indicators, such as correlation or causality, to infer interaction directly from the data \cite{SA2019}. This solution is based on time and frequency domain coherency relation between dynamics observed at different PMUs, which is backed by long-term industrial experience. However, there are still several practical challenges to achieving this goal: 1) Performing interaction learning and event identification as two separate stages would diminish the accuracy of event classification. 2) Most previous works require prior information on event location and system topology that is often not available to researchers due to privacy protection. For example, we are granted access to a dataset consisting of tens of PMUs with a time span of two consecutive years without disclosure of the grid topology. 3) Existing machine learning-based models that utilize multiple PMU data streams as input can lead to high model complexity, which makes their practical implementation costly. 

Another fundamental challenge for data-driven event detection and identification is the scarcity of real-world PMU data. Most data-driven models use a small amount of PMU data with limited labeled events or synthetic data. For example, in the study of the disturbance files at Public Service Company of New Mexico (PNM), only 97 events were labeled in the log-book, which are too few for training and testing a realistic event classifier \cite{SB2017}. In \cite{WWK2020}, hundreds of labeled frequency events from the FNET/GridEye system were used to train a deep learning-based frequency event detector. Generally speaking, small-scale datasets often do not cover enough scenarios and are too few to train and test a reliable event classifier realistically.

To address these challenges and the shortcomings in previous literature, we propose a novel graphical method that can integrate the interactive relationships of different PMUs to perform real-time event classification without requiring any knowledge of the system model/topology. Overall, we develop a deep learning-based model and train it with historical PMU data with the corresponding power system event labels. When the training process completes, the fitted model can be used as an online classifier to inform system operators of the types of system events using multiple PMU measurements. The uniqueness of the proposed method is the \emph{simultaneous optimization} of interaction graph inference, feature engineering, and event identification tasks, which can effectively mitigate the uncertainty of individual PMU data and improve the performance of the event classifier. To achieve this, spatial-based graph neural networks (GNNs) are integrated with an autoencoder architecture. In the encoder, for each labeled event, the latent relationship representing the probability of the existence of an edge between a pair of PMUs is estimated using a graph representation algorithm known as the deep relational network \cite{kipf}. Based on the latent graph relationship, a multi-layer graph structure is obtained via a deterministic graph sampling strategy. In the decoder, to efficiently construct event features based on the patterns of different event types, we propose an innovative dilated inception approach for extracting PMU data features. This method consists of multiple dilated convolutional layers with different dilation rates in a parallel manner, which can automatically capture multi-scale data features with limited parameters. By combining the interaction graph and data features, the graphical event classification can be performed. It should be noted that the proposed method is fine-tuned on our dataset to construct an end-to-end mapping relationship between PMU data features and event types predefined by data providers in this work. However, the proposed methodology is general. It can be used to perform various power event classification tasks (e.g., IEEE 1159 classification) when sufficient real event labels are available. The main contributions of this paper can be summarized as follows:
\begin{itemize}
\item The proposed method learns latent interaction graph jointly with feature engineering and event identification model, thus improving the accuracy of the graph learning and ensuring seamless integration between the learned interactions and event identification. 
\item The proposed event identification method integrates the spatial correlations of different PMUs fully in a data-driven manner, rather than assuming much a prior model knowledge, such as physical topology and event location. 
\item Instead of generating a single statistical graph to represent the pair-wise relationships among PMUs in different events, our approach generates different graphs for different power system events, thus dealing with uncertainty in the location and type of events.
\item The proposed model has been developed and tested based on a two-year real-world PMU dataset collected from the entire Western Interconnection in the U.S. The large number of real event labels contained in this dataset provides a good foundation for developing an efficient and practical event identification model. 
\end{itemize}

The rest of this paper is constructed as follows: Section \ref{overall} introduces the available PMU dataset and data pre-processing. In Section \ref{graph}, data-driven interactive relationship inference and graphical event classification are described. The numerical results are analyzed in Section \ref{result}. Section \ref{conclusion} presents research conclusions.

\section{Data Description and Pre-Processing}\label{overall}

\begin{figure}[tbp]
	\centering
	\includegraphics[width=3.5in]{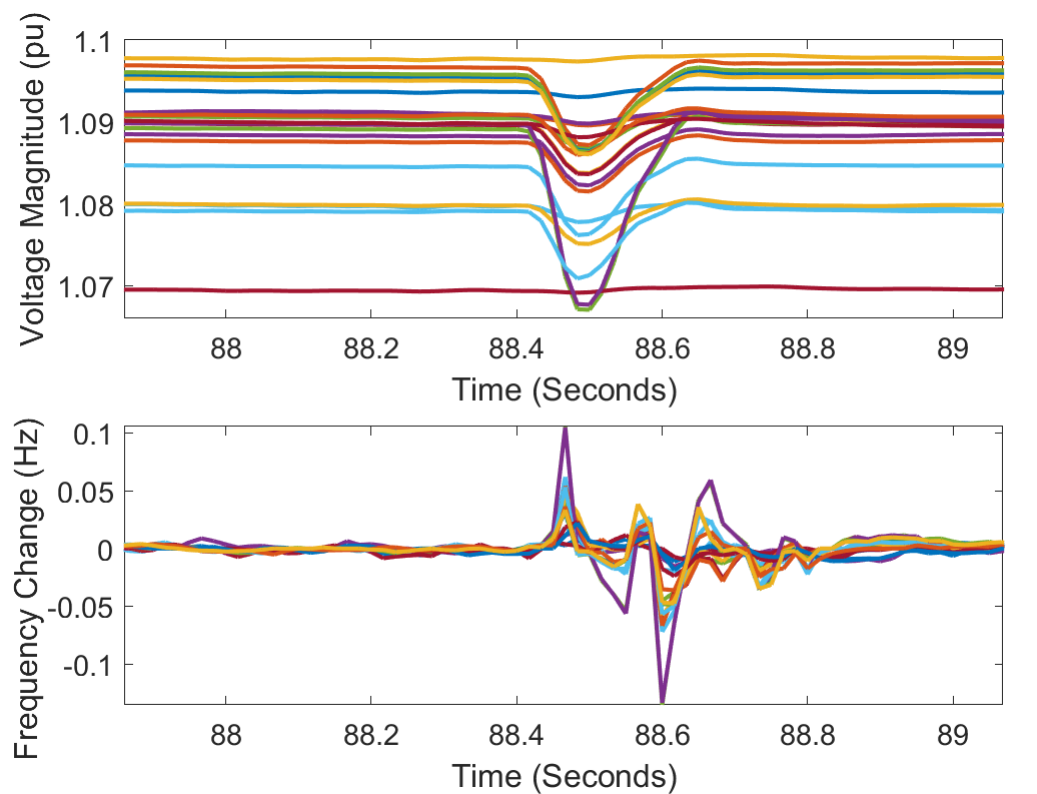}
	\caption{Plots of multiple PMU data for a real-life power event.}
	\label{fig:data_example}
\end{figure}

The proposed method is motivated by insights from real PMU data. The available data is obtained from 440 PMUs installed across three U.S. transmission interconnections, including the Texas, Western, and Eastern Interconnection\footnote{The dataset is stored as Parquet form and includes around two years of measurements, from 2016 to 2017. We have utilized Python and MATLAB to read and analyze the whole dataset, which is larger than 20 TB (around 670 billion data samples).}. The rates of sampling are 30 and 60 frames per second, and the measured variables include voltage and current phasor, system frequency, rate of change of frequency, and PMU status flag. For convenience, let A, B, and C denote the three interconnections hereinafter. Fig. \ref{fig:data_example} shows the voltage magnitude values and frequency variations of all PMUs in interconnection B for a specific event. Based on this figure, it is clear that all PMUs in an area have captured the event. However, even though the nature of the variations in PMU data will be similar (i.e., event patterns and start timestamps are almost the same), the amount of variations will be different \cite{SB2017}. Further, as demonstrated in the figure, several PMUs show negligible event features, which should be excluded from the inputs to the event classification model. To achieve this, one simple solution is to select the PMU that shows the biggest impact based on context information or specific metrics \cite{MB2016}. However, context information may be unavailable \textit{a prior} and metrics are hard to calculate in real time. Thus, in this work, we propose a more natural solution that utilizes data from all PMUs as input to the model and automatically selects the suitable PMUs and the associated data by discovering the interaction graphs. 
\begin{table}[tbp]
\caption{Statistical Summary of Three Interconnections.}
\centering\renewcommand\arraystretch{0.75}
\setlength{\tabcolsep}{0.2mm}{
\begin{tabular}{p{3.5cm}p{1.5cm}p{1.5cm}p{1.5cm}}
\hline\hline\\[-5pt]
 & A & B & C\\[0pt]
\hline\\[-5pt]
Start Time & 07/21/2018 & 01/01/2016 & 01/01/2016\\[2pt] 
End Time & 08/24/2019 & 12/31/2017 & 12/31/2017\\[2pt] 
Data Size & 3 TB & 5 TB & 12 TB\\[2pt]
Number of PMUs & 215 & 43 & 188\\[2pt]
Sample Rates [frames/s] & 30  & 30/60 & 30 \\[2pt]
Total Number of Events & 29 & 4854 & 1884\\[2pt]
Number of Unlabeled Events & 0 & 0 & 634\\[2pt]
Resolution of Event Record & Daily & Minute & Minute\\[2pt]
\hline\hline
\end{tabular}}
\label{table:stat}
\end{table}

Apart from PMU measurements, real event labels are needed to provide the ground truth for developing a practical PMU-based event identifier. In this work, a total of 6,767 event labels, consisting of 6,133 known events and 634 unknown events (where the event type entry is empty or unspecified), are utilized to extract the event data. Each event label includes the interconnection number, start timestamp, end timestamp, event type, event cause as well as event description. The timestamps of these event labels are determined by SCADA's outage alarm reception time in the control room. Also, the types of events have been verified with the corresponding protection relay records, ensuring a high level of confidence in the event labels. It should be noted that the definition of each event type is entirely up to the data provider. The detailed detection criteria for all types of events are unavailable for us due to the protection of sensitive information.

To prevent erroneous event detection due to data quality issues (i.e., bad data, dropouts, communication issues, and time errors), the available PMU dataset is initially passed through data pre-processing. Heterogeneous data quality issues are classified based on PMU status flag information. Following IEEE C37.118.2-2011 standard, when the value of the status flag is 0 in decimal format, data can be used properly; otherwise, data should be removed due to various PMU malfunctions. Also, we have utilized the engineering intuitions to design several simple threshold-based methods for further detecting the data quality problems not identified by PMU, such as out-of-range problem. Then, based on our data quality assessment, when a consecutive data quality issue occurs, the data is excluded from our study because it is hard to provide high accuracy data imputation for these consecutive bad data points. The remaining missing/bad data are filled and corrected by interpolation. In this work, an analysis window with length $T$ is utilized to extract event samples. The value of $T$ is assigned as 2-second based on previous works and observations of real PMU data \cite{SB2017,WG2011}. When the analysis window is large, the event classification model may suffer from the curse of dimensionality, thus resulting in serious overfitting problems. Also, as the input dimensionality increases, the computational complexity of the data-driven event classification model grows significantly. This will impact the real-time application of the model. Hence, the analysis window does not need to cover all event data, but needs to provide sufficient event features for identifying event types. Considering that the resolution of available event logs is in the order of minutes, we have used a statistical method to reach a finer scale \cite{yuanyx}. When the resolution of event logs is in the order of seconds, this statistical algorithm can be bypassed. Given that the available PMU dataset is more than 20TB, we have extracted post-event data for efficient event classification model development and testing based on the start timestamps of historical events recorded in the event log. It should be noted that we do not use all available data for model training due to the risk of data imbalance problems\footnote{The data imbalance problem refers to the uneven distribution of the number of observations in each category. In this work, the size of the post-event data is much smaller compared to the data in normal conditions. After training a supervised classification model using this dataset, the model always tends to classify the data points as normal operations to optimal classification accuracy.}. After data extraction, the time-series PMU data is converted into image-like data by applying a Markov-based feature reconstruction method from our previous work \cite{yuanyx}. To simulate the real situation faced by system operators, any manual modification to the event labels is avoided in this work. Even though the structure of the proposed model is \emph{fine-tuned} on our dataset, the methodology is general and can be applied to any PMU datasets after some fine-tuning procedures. This is true for any data-driven solution.

 \begin{figure*}[tbp]
	\centering
	\includegraphics[width=2\columnwidth]{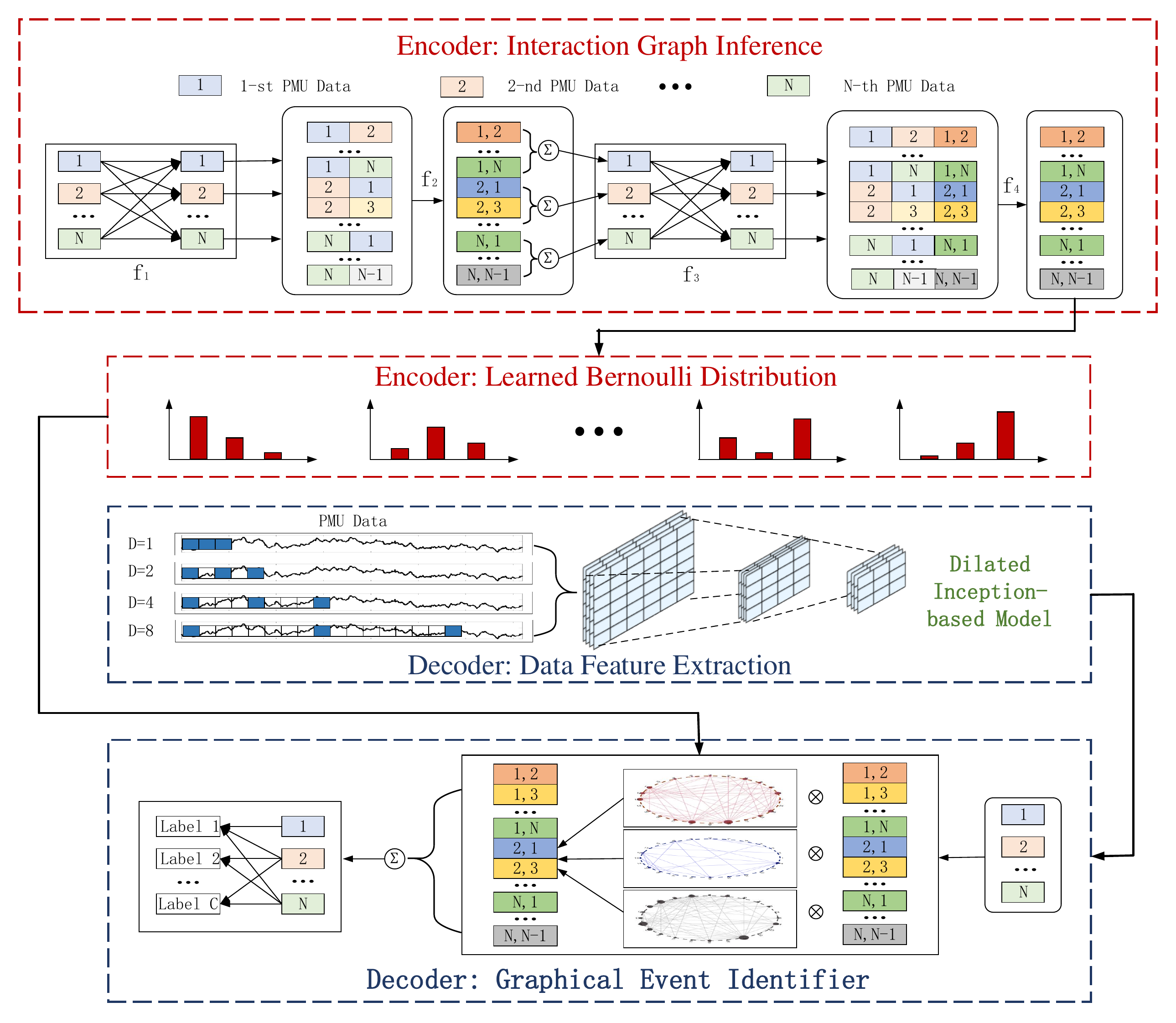}
	\caption{Overall structure of the proposed method.}
	\label{fig:main}
	\vspace{-3mm}
\end{figure*}
 \begin{figure}[tbp]
	\centering
	\includegraphics[width=3.5in]{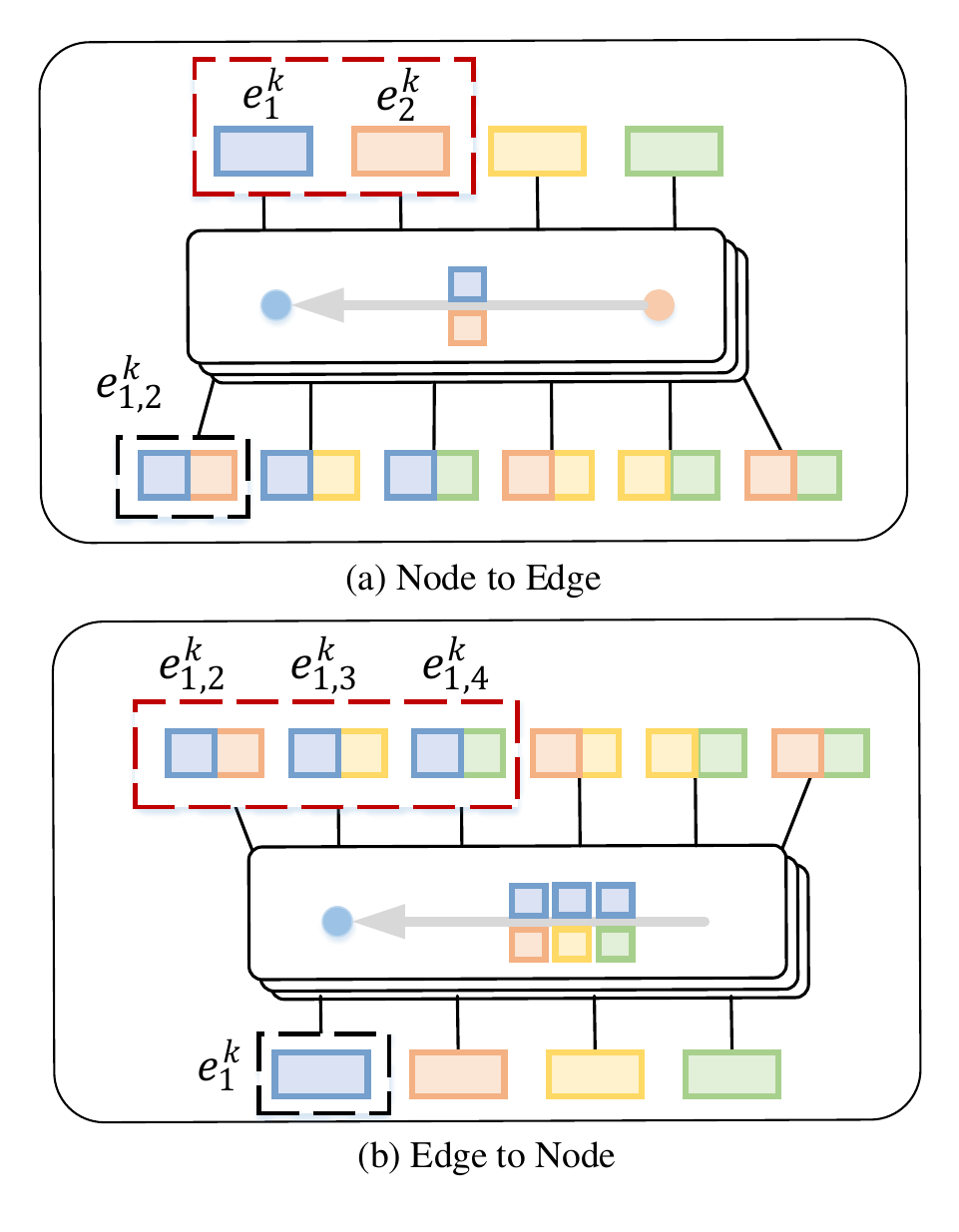}
	\caption{Interactive relationship inference procedure by using the node-to-edge and edge-to-node operations. }
	\label{fig:graph}
	\vspace{-2mm}
\end{figure}

\section{Graphical PMU-Based Event classification}\label{graph}
In this section, we lay out the proposed graphical event classification method. To help the reader understand our model, we first briefly revisit the concepts and properties of GNN, and then describe our method in detail.

Many real-world problems involve data that can be represented as a graph whose vertices and edges correspond to sets of entities and their relationships, respectively. Given that usual deep learning techniques are not applicable\footnote{Convolutional neural networks are well-developed for grid-structured inputs. Recurrent neural networks are well-defined only over sequence data.}, these problems have motivated the development of a class of neural networks for processing data represented by graph data structures, called GNNs. The key idea of GNN is to generate a representation of nodes, which actually depends on the structure of the graph, as well as any available feature information. According to existing studies \cite{GNN_2}, GNNs can be broadly categorized into spatial-based and spectral-based approaches. In general, spectral-based GNNs use eigendecompositions of the graph Laplacian to produce a generalization of spatial convolutions to graph, while providing access to information over short and long spatiotemporal scales simultaneously \cite{GNN_1}. In comparison, spatial-based GNNs involve a form of neural message-passing that propagates information over the graph by a local diffusion process \cite{GNN_3}. The proposed method falls into this categorization. 

In this work, spatial-based GNNs are combined with autoencoder to perform interaction learning and event classification jointly in an unsupervised way. Specifically, the encoder adopts spatial-based GNNs that act on the fully connected graph with multiple rounds of message passing and infer the potential interaction distribution based on all PMU measurements. The decoder uses another spatial-based GNN to identify event types based on PMU features and constructed graphs. The overall model is schematically described in Fig. \ref{fig:main}. Our work follows the line of research that learns to infer relational graphs while learning the dynamics from observational data \cite{kipf}\cite{LDS2019}. Unlike previous methods that focus on data prediction, the proposed method is capable of extracting multi-scale event features and performing accurate power system event classification. Moreover, since the interactions among different PMUs are impacted by the event location, our approach produces one graph structure for each event rather than a single statistics-based graph. Compared with existing bilevel optimization-based graph learning approach \cite{LDS2019}, the graph structure in our model is parameterized by neural networks rather than being treated as a parameter, thus significantly reducing the computational burden of data-driven interaction graph inference. In addition, the online computational cost of the proposed learning-based method is much lower than the optimization-based method, thanks to the neural network implementation. In the following, we describe the proposed model in detail.

\subsection{Interaction Graph Inference and Sampling}

Let us first settle the notations. In this work, each PMU and the corresponding data (i.e., voltage magnitude value) can be considered as a \emph{node} and an \emph{initial node feature}. Initial node features consist of $\{\mathbbm{V},\mathbbm{L}\}$, where $\mathbbm{V}:=\{v_{1},...,v_{h}\}$ is the voltage magnitude set from PMUs, $\mathbbm{L}:=\{l_{1},...,l_{h}\}$ is the corresponding event label set from the event logs, and $h$ is the total number of events. Specifically, $v_{i}\in\mathbbm{R}^{N\times T}$ is a set of voltage magnitude collected from $N$ PMUs during event $i$ within time windows with length $T$. Note that all PMU data in the same interconnection for a specific event are considered as one data sample in this work.

The goal of the encoder is to compute the latent relationship $\mathbbm{E}_{i,j}:=\{e_{i,j}^1,...,e_{i,j}^N\}$, where $e_{i,j}$ represents the probability of edge existence between PMUs $i$ and $j$. To achieve this, we utilize deep neural relational inference to pass local information \cite{GJ2017}:  
\begin{equation}
\label{eq:v_e_2}
e_{i,j}^k = f_e^k([e_i^k,e_j^k,x_{(i,j)}])
\end{equation}
\begin{equation}
\label{eq:e_v_1}
e_{i}^{k+1} = f_n^k([\sum_{i\in N_j} e_{i,j}^k, x_j]) 
\end{equation}
where, $e_{i}^{k}$ is the feature of node $i$ in layer $k$, $e_{i,j}^k$ is the feature of edge connecting nodes $i$ and $j$, $N_j$ is the set of edges connecting node $j$. $x_i$ and $x_{(i,j)}$ summarize initial nodes and edge features, respectively, and $[\cdot,\cdot]$ denotes the concatenation operation. The functions $f_e$ and $f_n$ refer to node- and edge-specific neural networks. The $f_e$ is mapped to compute per-edge updates. For example, for PMU 1 and 2, $e_{1,2}^k$ is calculated based on the features of PMU $1$ and $2$, $\{e_1^k,e_2^k\}$, as described in Fig. \ref{fig:graph}(a). The $f_n$ is utilized to compute per-node updates across all nodes. $\sum_{i\in N_j} e_{i,j}^k$ is obtained by aggregation of edge features from edges that are connected to node $i$, as shown in Fig. \ref{fig:graph}(b). Since we do not assume any a prior knowledge of the underlying PMU-based interaction graph, this operation is used on the fully connected graph (without self-loops). Note that if the operator has some knowledge on the latent/physical connections of PMUs, this fully connected graph can be easily replaced by a prior knowledge-based graph. For example, a Markovian influence graph formed from utility outage data is able to describe the temporal relationship between the disturbance dynamics of various PMUs \cite{kai}. Eqs. \eqref{eq:v_e_2} and \eqref{eq:e_v_1} allow for model combinations that represent node-to-edge/edge-to-node mappings through multiple rounds of message-passing \cite{WMK2019}. In this work, the encoder includes the following four steps to infer $\mathbbm{E}_{i,j}$:
\begin{equation}
\label{eq:encode_1}
e_i^1 = f_1(v_{i})
\end{equation}
\begin{equation}
\label{eq:encode_2}
\rm{Node}\rightarrow\rm{Edge:}\ e_{i,j}^1 = f_e^1([e_i^1,e_j^1])
\end{equation}
\begin{equation}
\label{eq:encode_3}
\rm{Edge}\rightarrow\rm{Node:}\ e_{i}^2 = f_n^1(\sum_{i\neq j} e_{(i,j)}^1) 
\end{equation}
\begin{equation}
\label{eq:encode_4}
\rm{Node}\rightarrow\rm{Edge:}\ e_{i,j}^2 = f_e^2([e_i^2,e_j^2])
\end{equation}
According to previous studies \cite{kipf}, two-layer fully connected neuron networks are utilized to model node- and edge-specific neural networks, which can be formulated as follows:
\begin{equation}
\label{eq:mlp}
f_1(v_i) = a(w_{f_1,0}^{(2)} + \sum_{i=1}^N w_{f_1,i}^{(2)} \cdot (a(w_{f_1,0}^{(1)} + \sum_{n=1}^N w_{f_1,n}^{(1)} \cdot v_n))
\end{equation}
where, $w_{f_1,0}, w_{f_1,1}, ..., w_{f_1,n}$ represent internal weights of $f_1$ and the exponential linear unit is used as the activation function $a$ in these networks. Compared to the commonly-used rectified linear unit, it has been shown that exponential linear units can achieve higher classification accuracy \cite{elu}. Also, to avoid internal covariate shift during training process, a batch normalization layer is added after the activation layer. As demonstrated concretely in \cite{SI2015}, the normalization is achieved by subtracting the batch mean and dividing by the batch standard deviation. It should be noted that the layer of the graph is determined by the number of output neurons in $f_e^2$, which is set as $3$ in this work.  

Using $\mathbbm{E}_{i,j}$, the interaction graph is obtained via a graph sampling technique. Here, we apply the following deterministic thresholding method:
\begin{equation}
\label{eq:deter_sample}
w_{i,j} = 
\begin{cases}
1 &\mathrm{if} \ \mathrm{sigmod}(e_{i,j})>r\\
0 &\mathrm{otherwise}
\end{cases} 
\end{equation}
where, $r$ is a user-defined threshold. The deterministic thresholding method encourages sparsity if $r$ gets closer to 1. Such a discrete graph, however, poses a challenge on differentiability. In other words, model parameters cannot be learned through backpropagation. To tackle this issue, we have utilized the Gumbel-Max trick, which provides an efficient way to draw samples from the categorical distribution \cite{gumble}. The detailed function is described as follows:
\begin{equation}
\label{eq:sample}
z = \rm{one}\_\rm{hot}(\argmaxA_m[g_m + \log{e_{i,j}^m}])
\end{equation}
where, $g_1,...,g_N$ are independent and identically distributed (i.i.d) samples drawn from the Gumbel distribution with 0 location and 1 scale parameter\footnote{Gumbel distribution with 0 location and 1 scale parameter can be sampled based on the inverse transform method: draw $u \sim $ standard uniform distribution and compute $g = -\log(-\log(u))$.}. Then, the softmax function is utilized as a differentiable approximation to $\argmaxA$:
\begin{equation}
\label{eq:sample_1}
z_{i,j}= \frac{\exp((log(e_{i,j}^m)+g_m)/\tau)}{\sum_{m=1}^N \exp((log(e_{i,j}^m)+g_m)/\tau)}
\end{equation}
where, $\tau$ is a smooth coefficient and is assigned as $0.5$ in this work. When $\tau \rightarrow{0}$, this approximated distribution converges to one-hot samples from $\mathbbm{E}_{i,j}$.

\begin{figure}[tbp]
	\centering
	\includegraphics[width=3.5in]{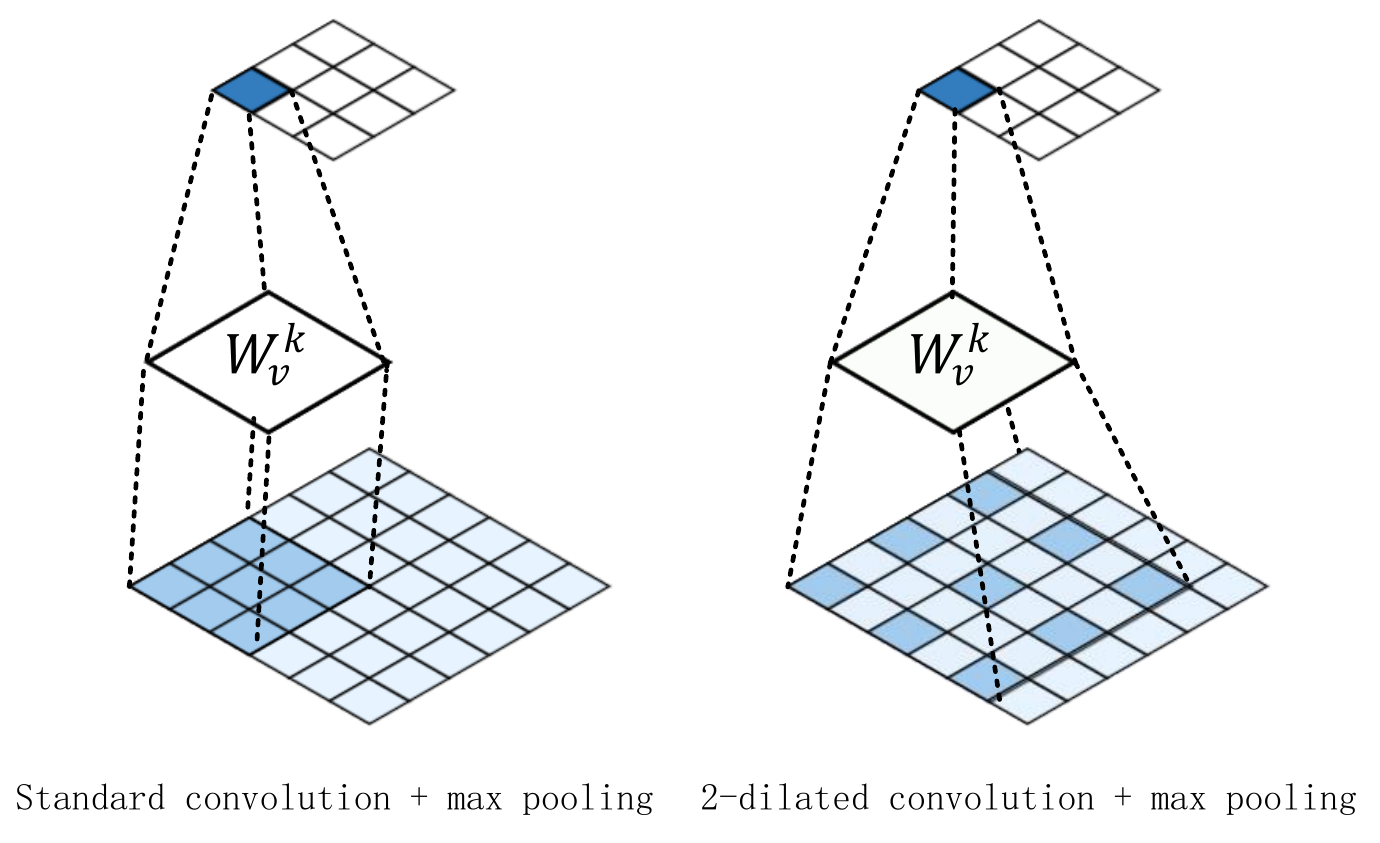}
	\caption{Illustrate of the two dilated convolutional layers and max-pooling layers.}
	\label{fig:dilated}
\end{figure}
 
\subsection{Feature Extraction and Event classification}

The goal of the decoder is to construct a mapping relationship between PMU data and event types. The basic idea is to fit a boundary in a high-dimensional space to separate data samples with different event types. To achieve superior classification performance in terms of both accuracy and efficiency, it is imperative to devise a good feature extractor. In our previous work \cite{yuanyx}, a Markov-based feature extractor was utilized to capture the multi-scale data features. However, this feature extractor has an exponential computational burden in terms of the dimensionality of the data samples, which is not appropriate in this work due to the extremely high-dimensional input. Hence, a new PMU-based feature extractor, dilated inception-based network, is proposed to capture multi-scale features effectively \cite{feature}. The proposed dilated inception-based network follows the line of the well-known convolutional layer for feature extraction and combination in a data-driven manner through fully end-to-end training \cite{Goodfellow2016}. To help the reader understand our model, we first review the standard convolutional layer and then describe the details of our method. The convolutional layer computes the convolutional operation, $\ast$, of the input using kernel filters to extract data feature maps, which can be mathematically formulated as follows \cite{auto2017}: 
\begin{equation}
\label{eq:convof}
\phi_{k}^\zeta = \sum_{u \in U}x_{k-1}^u\ast W_{k}^\zeta+b_{k}^\zeta
\end{equation}
where, $\phi_{k}^\zeta$ is the latent representation of the $\zeta$'th feature map of the $k$'th layer; $x_{k-1}^u$ is the $u$'th feature map of the previous layer and $U$ is the total number of feature maps; $W_{k}^\zeta$ and $b_{k}^\zeta$ are the kernel filter and the bias of the $\zeta$'th feature map of the $k$'th layer, respectively. In this work, $x_{k-1}^u\ast W_{k}^\zeta$ can be rewritten as follows:
\begin{equation}
\label{eq:convo}
(x_{k-1}^u\ast W_{k}^\zeta)(i,j)=\sum_{\delta_{i}=0}^{U-1}\sum_{\delta_{j}=0}^{U-1}x_{k-1}^u(i-\delta_{i},j-\delta_{j})W_{k}^\zeta(i,j) 
\end{equation}
where, $i$ and $j$ are the row and column indices of the PMU-based Markov matrix. Hence, the convolutional layer operates in a sliding-window way to output the feature maps. For each convolutional layer, the size of the output feature map is  $\phi_{k}^\zeta \in \mathbbm{R}^{(p-q+1) \times (p-q+1)}$, where $x_{k-1}^u$ and $W_{k}^\zeta$ are $p \times p$ and $q \times q$ matrices, respectively.

The main idea of dilated convolution is to insert zeros between two consecutive features in the convolutional kernel, which significantly increases the receptive field\footnote{In the context of deep learning, the receptive field is the region in the input space where the features are generated.}. In general, the dilated convolution operation is defined as:
\begin{equation}
\label{eq:dilated}
y_k^u(i) = \sum_l x_{k-1}^u(i+r\cdot l)\ast W_k^\zeta(l)
\end{equation}
where, $r$ is a dilation factor. For a $n \times n$ dilated kernel filter, the actual size of the receptive field is $n_d \times n_d$, where $n_d = n+(n-1)\cdot (r-1)$. This indicates that higher $r$ can capture more slowly-varying features over a larger temporal window. When $r$ equals 1, the standard discrete convolution is equivalent to the 1-dilated convolution. A comparison between standard convolution and dilated convolution is described in Fig. \ref{fig:dilated}. It is clear that a dilated $3 \times 3$ convolution kernel with $r=2$ has a similar receptive field with a standard $5 \times 5$ convolution kernel. To achieve multi-scale feature extraction, four dilated convolutions with various dilation rates are used in a parallel manner. The values of dilation rates are determined based on the validation set. After each dilated convolution layer, a max-pooling layer is added to summarize the feature maps. Max-pooling can be considered as a sample-based discretization procedure based on the feature map from the previous layer. This is achieved by dividing the input matrix into $N_{\rm{out}}^2$ pooling regions $P_{i,j}$ and selecting the maximum value \cite{BG2014}:
\begin{equation}
\label{eq:pooling}
P_{i,j} \subset \{1,2,...,N_{\rm{in}}\}^2, \forall (i,j)\in\{1,2,...,N_{\rm{out}}\}^2.
\end{equation}
In this work, a $4 \times 4$ max-pooling is used. Thus, $N_{\rm{in}}=4N_{\rm{out}}$ and $P_{i,j}=\{4i-1,4i\}\times \{4j-1,4j\}$. As a result, a feature matrix is obtained: $U_i = \{u_{i,1},...,u_{i,T'}\}$, where $T'$ is the reduced data length.

When the PMU features are obtained, GNN is utilized to perform the event classification task \cite{brain}. Compared to previous machine learning-based methods that use only data features as model input, our event identifier combines data features and interaction graph. To achieve that, a node-to-edge operation is performed on the extracted edge feature. Then, the obtained graph structure is combined with edge features using the element-wise multiplication ($\otimes$). The process can be formulated as follows:
\begin{equation}
\label{eq:GNN_dec}
h_{i,t} = \sum_{i\neq j}\sum_{k=1}^K w_{i,j}\cdot g_1([u_{i,t},u_{j,t}])
\end{equation}
Similar to the encoder, the node-based function $g_1$ is represented by a two-layer fully connected network that includes rectified linear units as the activation function, which can be formulated as follows:
\begin{equation}
\begin{split}
\label{eq:mlp_1}
g_1([u_{i,t},u_{j,t}]) &= max(0, w_{g_1,0}^{(2)} + \sum_{i=1}^N w_{g_1,i}^{(2)} \cdot \\
&max(0, w_{g_1,0}^{(1)} + \sum_{n=1}^N w_{g_1,n}^{(1)} \cdot [u_{i,t},u_{j,t}]))
\end{split}
\end{equation}
The event classifier is achieved by adding a two-layer fully connected network after vectorization, as follows:
\begin{equation}
\label{eq:GNN_dec_2}
\hat{l_i} = g_2([vec(U_i),vec(H_i)])
\end{equation}
where, $H_i = [h_{i,1},...,h_{i,T}]$. In this fully connected network, the softmax activation function is applied to normalize the output to a probability distribution over estimated event types:
\begin{equation}
\begin{split}
\label{eq:mlp_2}
g_2([vec&(U_i),vec(H_i)]) = softmax(w_{g_2,0}^{(2)} + \sum_{i=1}^N w_{g_2,i}^{(2)} \cdot \\
&max(0, w_{g_2,0}^{(1)} + \sum_{n=1}^N w_{g_2,n}^{(1)} \cdot [vec(U_i),vec(H_i)])
\end{split}
\end{equation}

\subsection{Hyperparameters Calibration}
Considering that the hyperparameters of all machine learning models (i.e., the number of layers and neurons, the dilation rate, the deterministic threshold) affect performance, the model has to be well-designed. The rationale behind the model design is to make a trade-off between model complexity and classification accuracy. Hence, we utilize the random search method to find the appropriate hyperparameter sets in this work \cite{randomsearch}. Basically, the value of the hyperparameter is chosen by "trial and error". It is hard to say that the selected hyperparameters are optimal, but these hyperparameters can provide good accuracy for the available real-world dataset with limited model complexity. Specific values of hyperparameters are listed in the numerical section. For model training, the adaptive moment estimation (Adam) algorithm with a learning rate of 0.001 is used to update the learning parameters of the proposed model \cite{adam}. Adam is an adaptive learning rate optimization for training deep neural networks. Based on the adaptive estimation of lower-order moments, Adam can compute individual adaptive learning rates for each parameter, which significantly increases the training speed \cite{adam}.


\subsection{Overfitting Mitigation Strategy}
The superior performance of deep learning models relies heavily on the availability of massive training data samples. Unlike our previous work that treated each PMU independently and enjoyed a high level of data redundancy\footnote{In our previous PMU-based event classification model, we have utilized the data of a single PMU to construct a training dataset, which is more than 200,000 data samples.}, the proposed graphical model is trained with the limited event-based data samples. Therefore, it is imperative to deal with the overfitting problem. To facilitate a better understanding, we first provide a simple explanation of the overfitting problem. Overfitting refers to a learning model that can only model the training data well. If a model suffers from an overfitting problem, the accuracy of the model for unseen data is questionable. Hence, three strategies are utilized to eliminate the overfitting problem in this work.

\textbf{Dropout:} Dropout is a commonly-used regularization method to prevent model overfitting \cite{dropout}. The basic idea of dropout is to randomly set the outgoing edges of hidden units to $0$ at each iteration of the training procedure. In this work, based on the calibration results, the dropout ratio that specifies the probability at which outputs of the layer are temporarily dropping out is set as 0.3. 

\textbf{Constraining model complexity:} 
As demonstrated in Fig. \ref{fig:main}, the proposed model possesses a relatively high model complexity compared to conventional classification models due to the presence of graph learning and multi-scale feature extractor. One natural way to reduce the risk of overfitting is to constrain model complexity \cite{Goodfellow2016}. To achieve this, the number of adaptive parameters (i.e., the number of hidden neurons in $f_1$, $f_e^1$, $f_n^1$, and $f_e^2$ functions) in the network is reduced. 

\begin{table}[tbp]
\caption{The structure of the graphical event classification model.}
\centering
\renewcommand{\arraystretch}{0.8}
\begin{tabular}{ccc}
\hline\hline
Panel & Type & Output Shape\\[2pt]
\hline
1/1 & 2-layer MLP & (16,24,256) \\[3pt]
1/2 & Batch normalization & (16,24,256) \\[3pt]
1/3 & Node-edge operation & (16,552,256) \\[3pt]
\hline
2/1 & 2-layer MLP & (16,552,256) \\[3pt]
2/2 & Batch normalization & (16,552,256) \\[3pt]
2/3 & Edge-node operation & (16,24,256) \\[3pt]
\hline
3/1 & 2-layer MLP & (16,24,256) \\[3pt]
3/2 & Batch normalization & (16,24,256) \\[3pt]
3/3 & Node-edge operation & (16,552,256)\\[3pt]
\hline
4/1 & 2-layer MLP & (16,552,256) \\[3pt]
4/2 & Batch normalization & (16,552,256) \\[3pt]
4/3 & fully connected layer & (16,552,3) \\[3pt]
\hline
5/1 & Dilated-inception model (4 parallel dconv1d) & (384,32,30) \\[3pt]
5/2 & Dilated-inception model (4 parallel dconv1d) & (384,32,7) \\[3pt]
5/3 & Dilated-inception model (4 parallel dconv1d) & (384,32,1) \\[3pt]
\hline
6/1 & fully connected layer & (16, 1, 552, 256) \\[3pt]
6/2 & Activation layer & (16, 1, 552, 256) \\[3pt]
6/3 & fully connected layer & (16, 1, 552, 256) \\[3pt]
6/4 & Activation layer & (16, 1, 552, 256) \\[3pt]
\hline
7/1 & fully connected layer & (16, 256) \\[3pt]
7/2 & Activation layer & (16, 256) \\[3pt]
7/3 & fully connected layer & (16, 5) \\[3pt]
\hline\hline
\end{tabular}
\label{table:1.1}
\end{table}

\textbf{Data augmentation:} Theoretically, one of the best options for alleviating overfitting is to get more training data. It is well-known that collecting enough power event data is hard and time-consuming, yet we still could easily increase the size of the training dataset and reduce the degree of data imbalance by leveraging data augmentation technology \cite{dataaug}. Here, we utilize a horizontally flipping method to obtain additional data samples. To eliminate the impact of the event location, in the data augmentation, we do the same procedure for all PMU signals in a given event. Moreover, the Gaussian noise with 0 mean and $0.04$ variance is added to these additional data samples. 

\subsection{Challenges of Imperfect PMU Data}
In actual grids, data quality issues, such as bad data, dropouts, and time error, arise frequently, and can easily impact any data-driven event classification solution, as described in the literature \cite{pmu_bad_data}. The rationale behind this is that the data qualify problems lead to the problem of imbalance in data dimensions. During the offline training process, data quality is solved by dropping data points. In the online process, one common solution is to perform data imputation methods (i.e., artificially generated data points based on data history) to eliminate the impact of missing and bad data on the proposed graphical event classification method. Also, our previous work, namely spatial pyramid pooling-aided method \cite{yuanyx}, can be easily integrated with the decoder of the proposed graphical model to eliminate the impact of missing and bad PMU data online. This SPP-aided mechanism can offer a unique advantage: the dimensionality of the test data can be different from that of the training data, which provides a fundamental solution to the online PMU data quality problem. More technical details can be found in \cite{yuanyx}.

\subsection{Application Challenges}
As detailed below, we discuss two application challenges:
\begin{itemize}
\item In actual grids, utilities may have incomplete event logs (i.e., the majority of events are unknown). It is well-known that collecting tremendous high-quality event labels is expensive. Most utilities may only have a limited number of labeled events. This lack of knowledge may reduce the accuracy and generalization of the proposed model. 
\item As a supervised learning-based model, the proposed method assumes that labeled events (i.e., record in event logs) and unseen events come from the same distribution. In other words, all event types need to be observed and registered in event logs. However, such an assumption may be difficult to hold in practice, among which one common case in actual grids is that unspecified event contains types that are never observed by system operators. When the features of unseen event types are intertwined with the features of recorded event types, such a class distribution mismatch problem can increase the difficulty of event identification. 
\end{itemize}

\section{Numerical Results}\label{result}
This section explores the practical effectiveness of our proposed graphical event classification model by using a real-world dataset. As detailed below, we test our model on PMU measurements and the related event logs of interconnection B. Interconnection B consists of approximately 136,000 miles of transmission lines and serves more than 80 million people in 14 states. The entire dataset includes about 4,800 event data samples, including line outages, transformer outages (XFMR), and frequency events, as well as 4,800 data samples under normal conditions. After data cleaning, the available dataset, including the PMU measurements and related event labels, is randomly divided into three separate subsets for training (70$\%$ of the total data), validation (15$\%$ of the total data), and testing (15$\%$ of the total data). Moreover, to make the model development procedure more rigorous so as to ensure that the proposed model has good reliability, we have applied a $k$-fold cross-validation strategy, where $k$ is selected as 5 in this work. Specifically, all data except the testing set is partitioned into $k$ disjoint folds and one of the $k$ folds is used as the validation set while using all remaining folds as the training set. This procedure is repeated until each of the $k$ folds has served for model validation. In other words, all data in the available dataset have been treated as unseen data for model development. When the training process completes, all data in the testing set is treated as unseen data to assess the final performance of our model.

\subsection{Performance of the Graphical Event classification}
The case study is conducted on a standard PC with an Intel(R) Xeon(R) CPU running at 4.10GHZ with 64.0GB of RAM and an Nvidia Geforce GTX 1080ti 11.0GB GPU. To help the reader understand each step of the proposed model, the detailed structure of the proposed PMU-based event identifier is presented in Table \ref{table:1.1}. In this table, we provide the type and output shape for each layer. As can be seen, our model mainly includes seven panels to achieve event classification using PMU data. More precisely, the encoder consists of the first four panels for interaction graph inference. The encoder includes the last three panels for data feature extraction and graphical neural networks. Depending on this model structure, the event classification performance of the proposed model is developed and evaluated on the training set and testing set, respectively. One shortcoming of the autoencoder architecture is the high computational complexity, especially for the training process. In our experiments, the training time is about 10 hours. However, given that the training procedure of our method is an offline process, the high computational cost of the training process does not impact the real-time performance of the proposed method. Based on 1440 testing samples, the average testing time for the proposed method is about 0.02 seconds due to the proposed parallel feature engineering. Consequently, in actual grids, when the input data arrives at the phasor data concentrator (PDC) from multiple PMUs, the proposed method can provide estimated results in roughly 200 ms, including the communication delays, which is much faster than heuristics-based methods. Without encoders, the average training and testing time of the dilated inception-based event classifier can be reduced to 3 hours and 0.013 seconds, respectively.



\begin{figure}[tbp]
	\centering
	\includegraphics[width=3.5in]{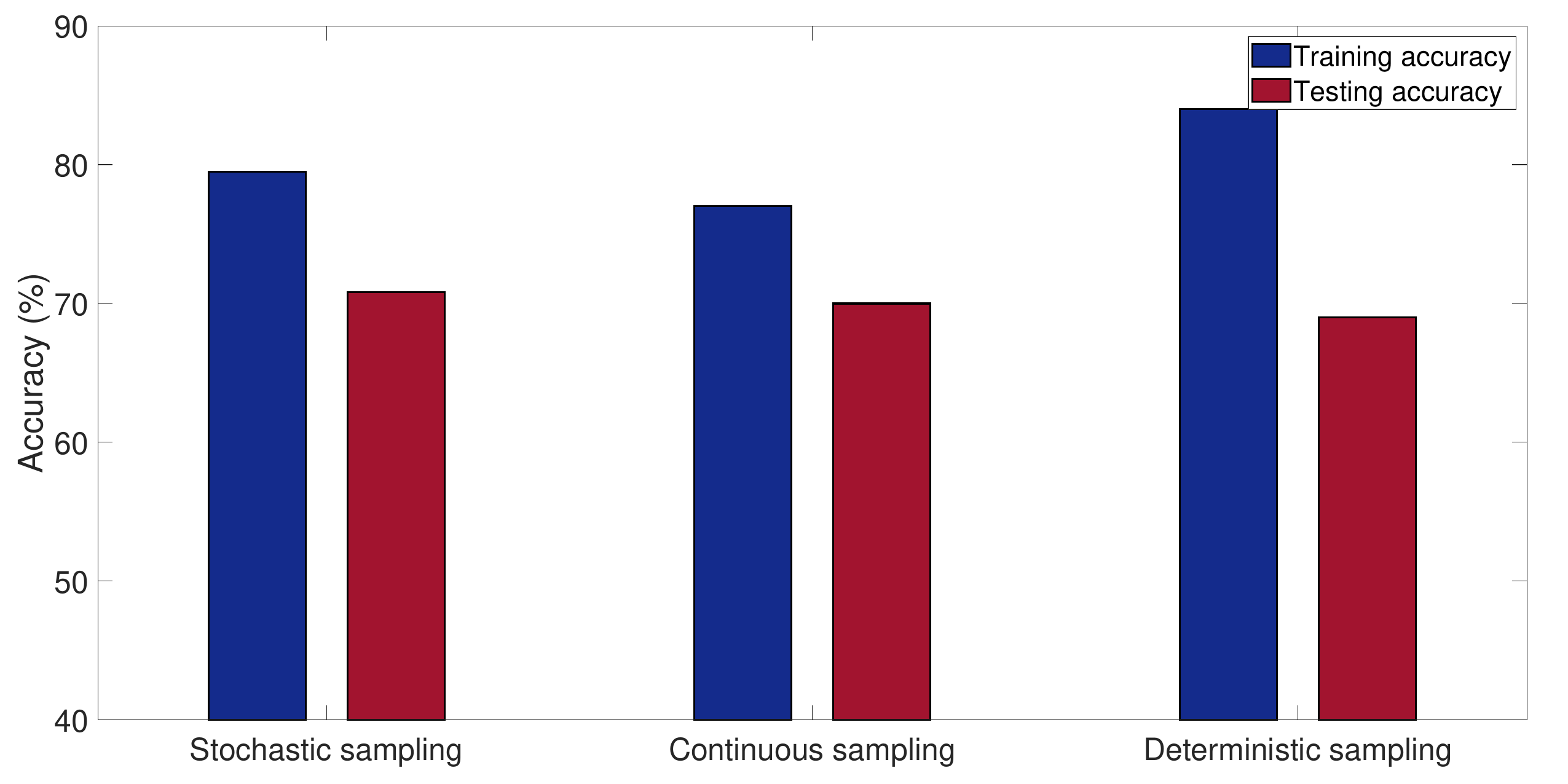}
	\caption{Comparison of three different graph sampling methods.}
	\label{fig:sample}
\end{figure}

\begin{figure}[tbp]
	\centering
	\includegraphics[width=3.5in]{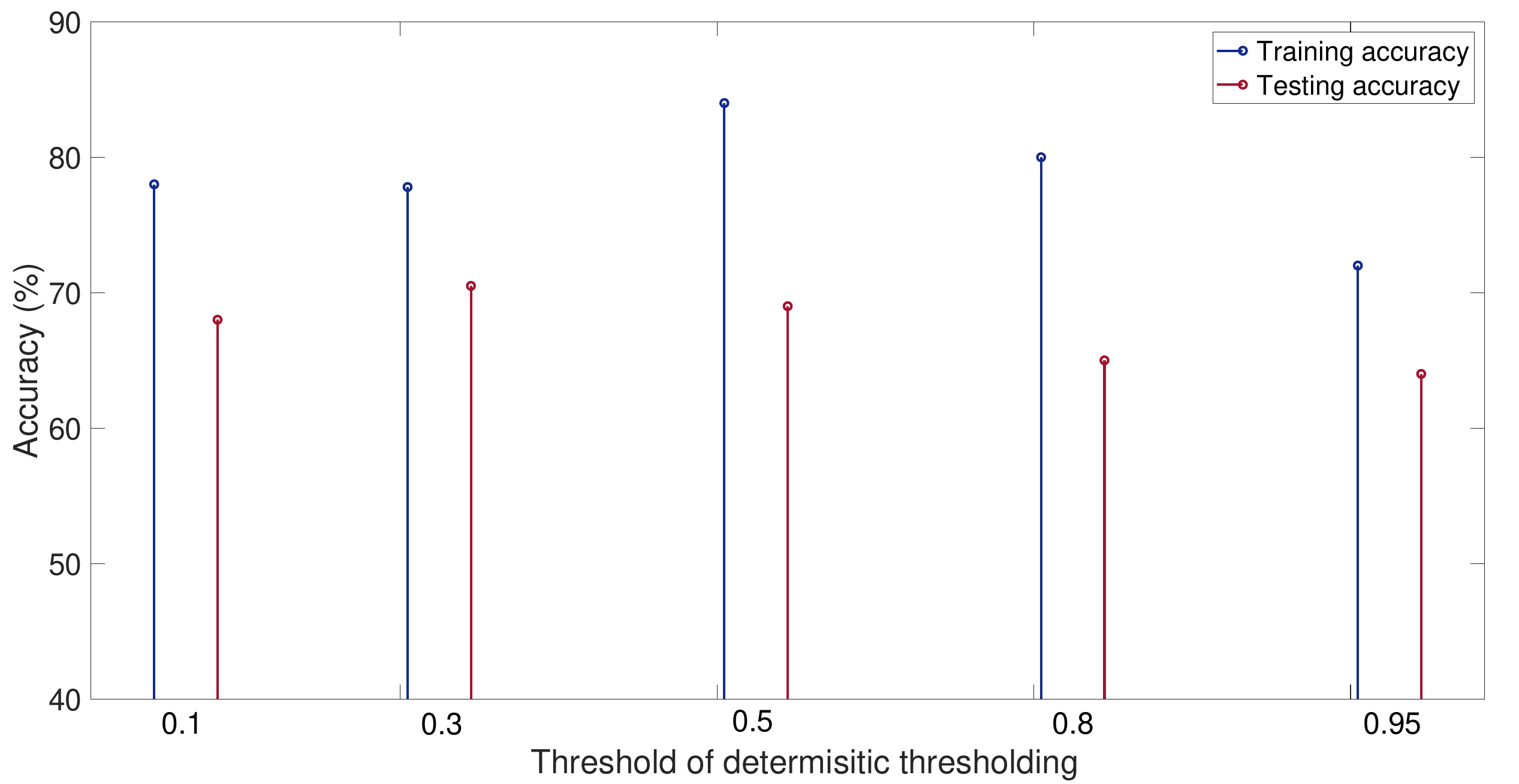}
	\caption{Sensitivity of event classification accuracy to the graph sparsity.}
	\label{fig:sparsity}
\end{figure}

The performance of the proposed method is evaluated by using real event logs recorded by utilities. First, we show the accuracy of our model under various graph sampling methods (i.e., stochastic sampling, continuous sampling, and deterministic thresholding) and feature extractors (i.e., standard convolutional layer and dilated inception network). Note that the following results are obtained by using the same overfitting strategy (dropout). As shown in Fig. \ref{fig:sample}, the training and testing accuracy values for the three graph sampling methods are $\{77\%,79.5\%,84\%\}$ and $\{70\%,70.8\%,69\%\}$, respectively. Based on this dataset, the deterministic thresholding method shows slightly better performance than other sampling methods. Moreover, Fig. \ref{fig:sparsity} is plotted to represent the sensitivity of the classification accuracy to the graph sparsity (i.e., the threshold of the deterministic thresholding method). As depicted in the figure, the performance of the proposed model can reach better accuracy with a moderate threshold value (around 0.5). Extremely high or low threshold values are inappropriate.

\begin{figure}[tbp]
	\centering
	\includegraphics[width=3.5in]{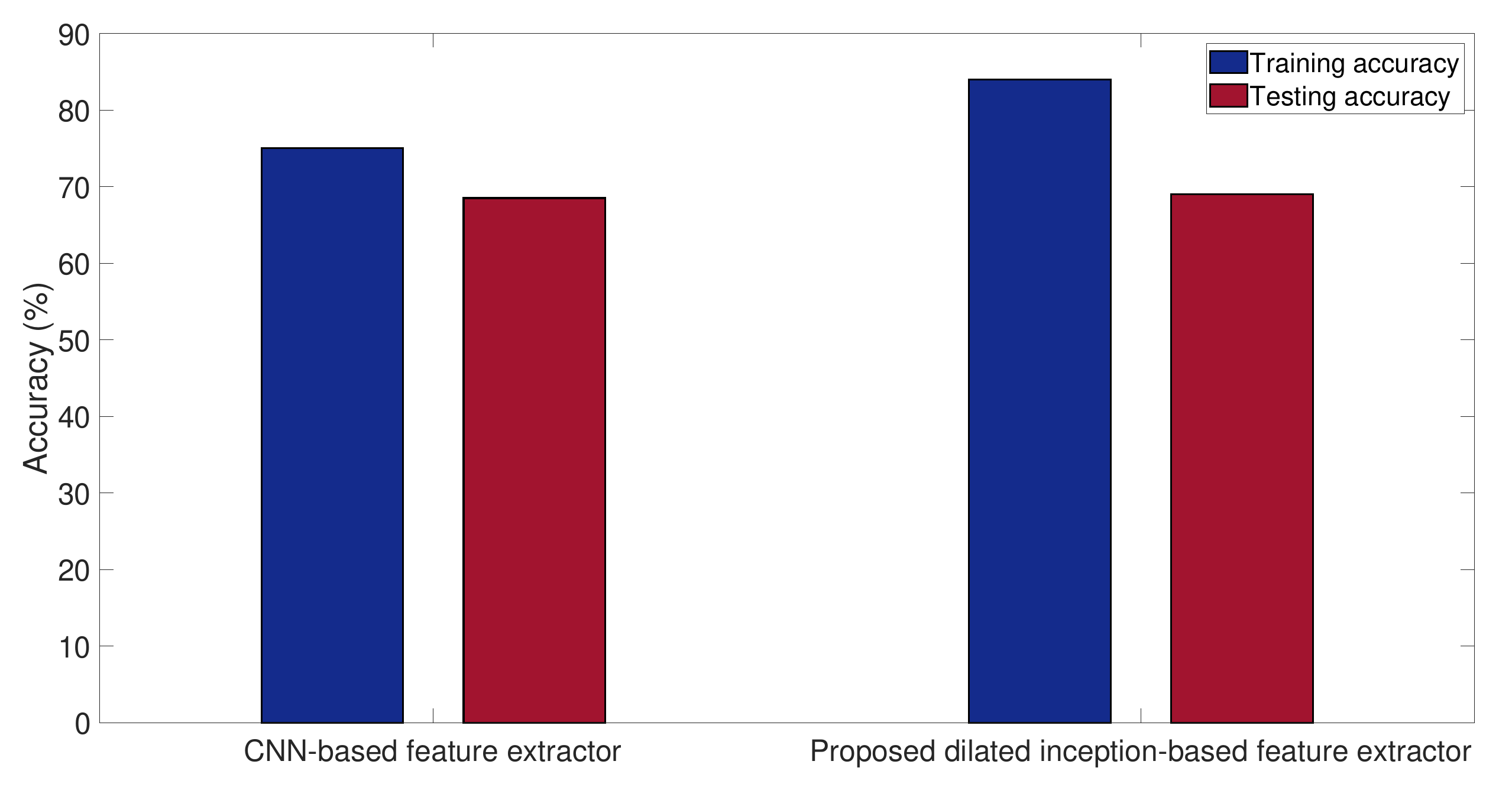}
	\caption{Comparison of CNN-based feature extractor and proposed dilated inception-based feature extractor.}
	\label{fig:feature_exactor}
\end{figure}

\begin{figure}[tbp]
	\centering
	\includegraphics[width=3.5in]{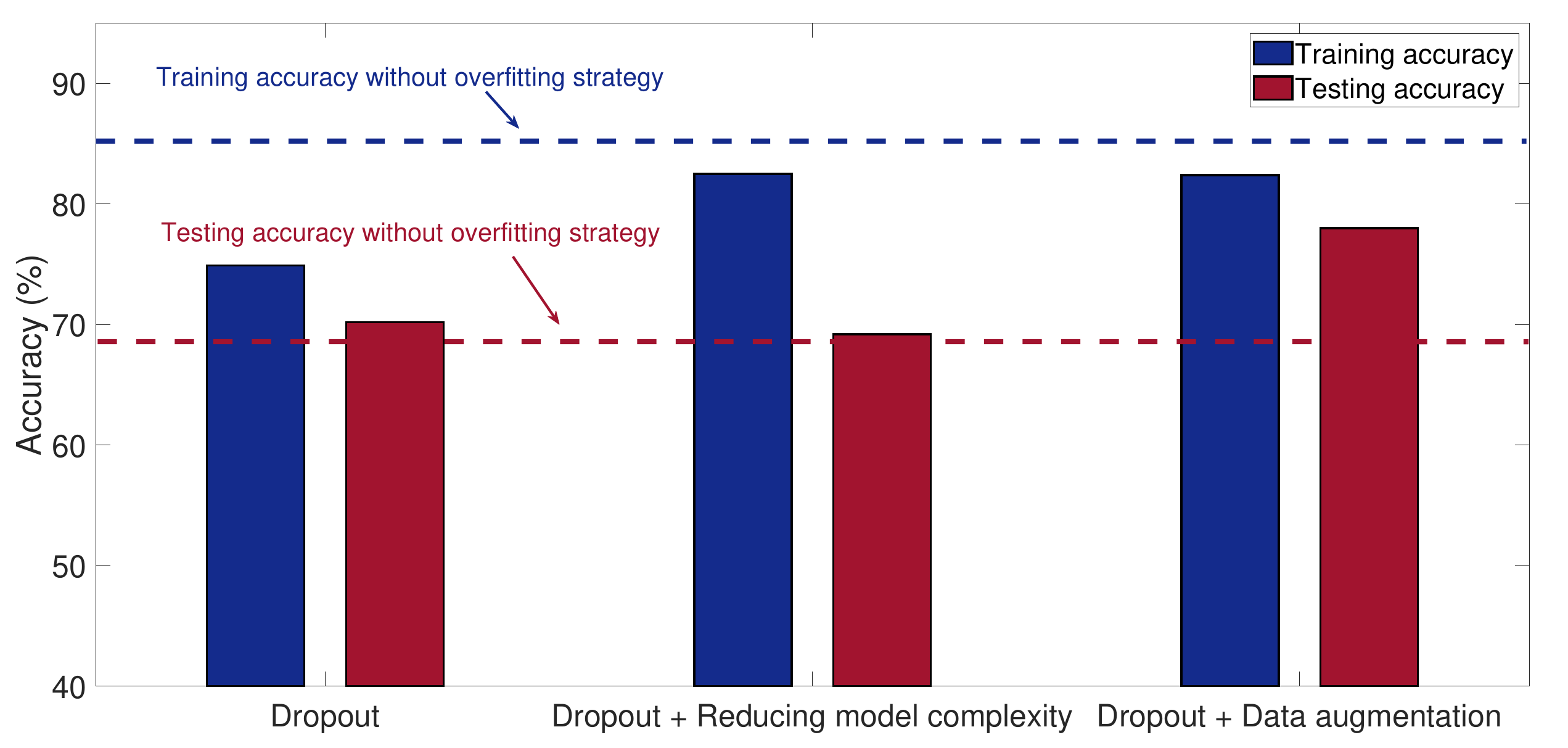}
	\caption{Comparison of three overfitting strategies.}
	\label{fig:overfit}
\end{figure}

\begin{figure}[tbp]
	\centering
	\includegraphics[width=3.5in]{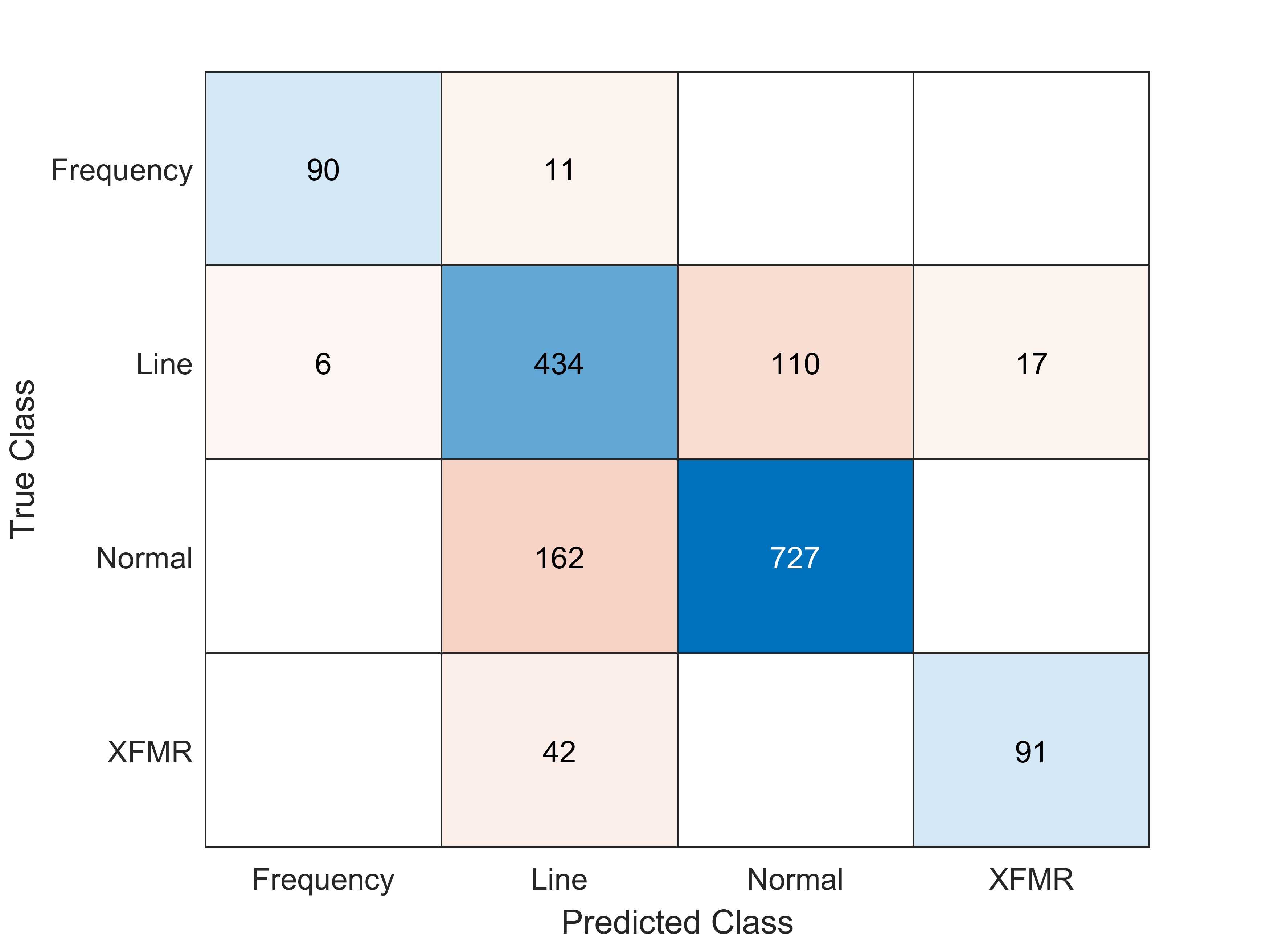}
	\caption{Confusion matrix for interconnection B using the proposed model.}
	\label{fig:conf}
\end{figure}

Then, two different feature extractors, namely the proposed dilated inception-based feature extractor and traditional CNN (including 3 convolutional and 2 max-pooling layers) are compared, as shown in Fig. \ref{fig:feature_exactor}. In this figure, the training and testing accuracy of the proposed dilated inception-based feature extractor, $\{84\%,69\%\}$, are higher than the values of the traditional CNN structure, $\{75\%,68.5\%\}$, which proves the enhancement of the multi-scale feature extractor. However, based on Fig. \ref{fig:sample}-\ref{fig:feature_exactor}, it is observed that the difference between the training and testing accuracy is not trivial. This indicates that the dropout strategy falls short of dealing with the overfitting problem in this case. Hence, we have combined two other strategies: constraining model complexity and data augmentation. The corresponding accuracy values are presented in Fig. \ref{fig:overfit}. As seen in the figure, the training accuracy decreases from around $84\%$ to around $82\%$. However, the testing accuracy is significantly higher compared to the previous cases. In this case, the combination of dropout and data augmentation has the best performance in reducing the overfitting risk: the training and testing accuracy are $\{82.4\%,78\%\}$. It is clear that the testing accuracy of the model will eventually achieve a similar level with the training accuracy if we can add more data samples. 

To show the performance of our method for different kinds of events, we have added a confusion matrix, as shown in Fig. \ref{fig:conf}. In this figure, the rows correspond to the estimated type and the columns correspond to the true type. The diagonal and off-diagonal cells correspond to events that are correctly and incorrectly classified, respectively. As seen in this figure, even though the available dataset is highly unbalanced, the proposed method still can identify most power system events, including line outages, XFMR outages, and frequency events. Moreover, except for accuracy, we have calculated precision, recall, and $F_1$ score to further show the performance of our method for each event type \cite{roc2006}. These indexes are determined as follows:
\begin{equation}
\label{eq:Prec}
Precision=\frac{TP}{TP+FP}
\end{equation}
\begin{equation}
\label{eq:recall}
Recall=\frac{TP}{TP+FN}
\end{equation}
\begin{equation}
\label{eq:f1}
F_1=\frac{(\beta^2+1)\cdot Prec\cdot Recall}{(\beta^2\cdot Prec+Recall)}
\end{equation}
where, TP is the true positive (i.e., an event is classified as line outage while its actual event type is also line outage), FP is the false positive (i.e., an event is classified as line outage while its actual event type is not line outage), FN is the false negative (i.e., an event is classified as other while its actual event type is line outage), and $\beta$ is the precision weight which is selected to be 1 in this work. The values of these indexes are presented in Table \ref{table:1.2}. 

Note that we are not surprised that the values of these indexes do not exceed 90\% on this dataset. In our opinion, there are two reasons that limit the accuracy of the proposed methodology. The first one is that the proposed method is based solely on a real-world PMU dataset. Unlike artificial datasets with clear and easy-to-see event patterns, real-world datasets suffer from noise and data quality issues, leading to degraded model performance. Meanwhile, we have applied the fully connected graph as the basic graph in the interaction graph inference process to avoid the assumption that the topology of the transmission system is known. This will increase the difficulty of latent relationship mining and therefore further impact the accuracy of the algorithm. The second one is that data augmentation operations can alter the data distribution during the training progress. This imposes a data distribution bias between the augmented data and the original data, which may reduce model performance. One of the best ways to deal with the overfitting problem in power event classification models is to simulate event samples based on the same transmission system, as described in \cite{SB2017}. Given that we currently do not have access to the topology of the interconnections and the spatial information of PMUs due to privacy protection, future work will be done to meet the gap once we acquire this information. 

\begin{table}[tbp]
\caption{Event classification analysis.}
\centering
\setlength{\tabcolsep}{6mm}{
\begin{tabular}{cccc}
\hline\hline
 & Precision & Recall & $F_1$ Score\\[3pt]
\hline
Normal & 0.8178 & 0.8686 & 0.8424\\[3pt]
Line & 0.7654 & 0.6687 & 0.7138\\[3pt]
XFMR & 0.6842 & 0.8426 & 0.7552\\[3pt]
Frequency & 0.8911 & 0.9375 & 0.9137\\[3pt]
\hline
\end{tabular}}
\label{table:1.2}
\end{table}

\begin{figure}[tbp]
	\centering
	\includegraphics[width=3.5in]{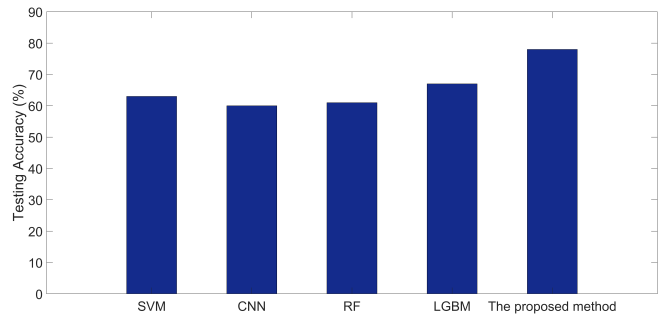}
	\caption{Comparison results of the proposed method and four existing event classification models.}
	\label{fig:compare_new}
\end{figure}

\begin{figure}[htbp]
\centering
\subfloat [Representative graph structure for all training data.]{
\includegraphics[width=2.7in]{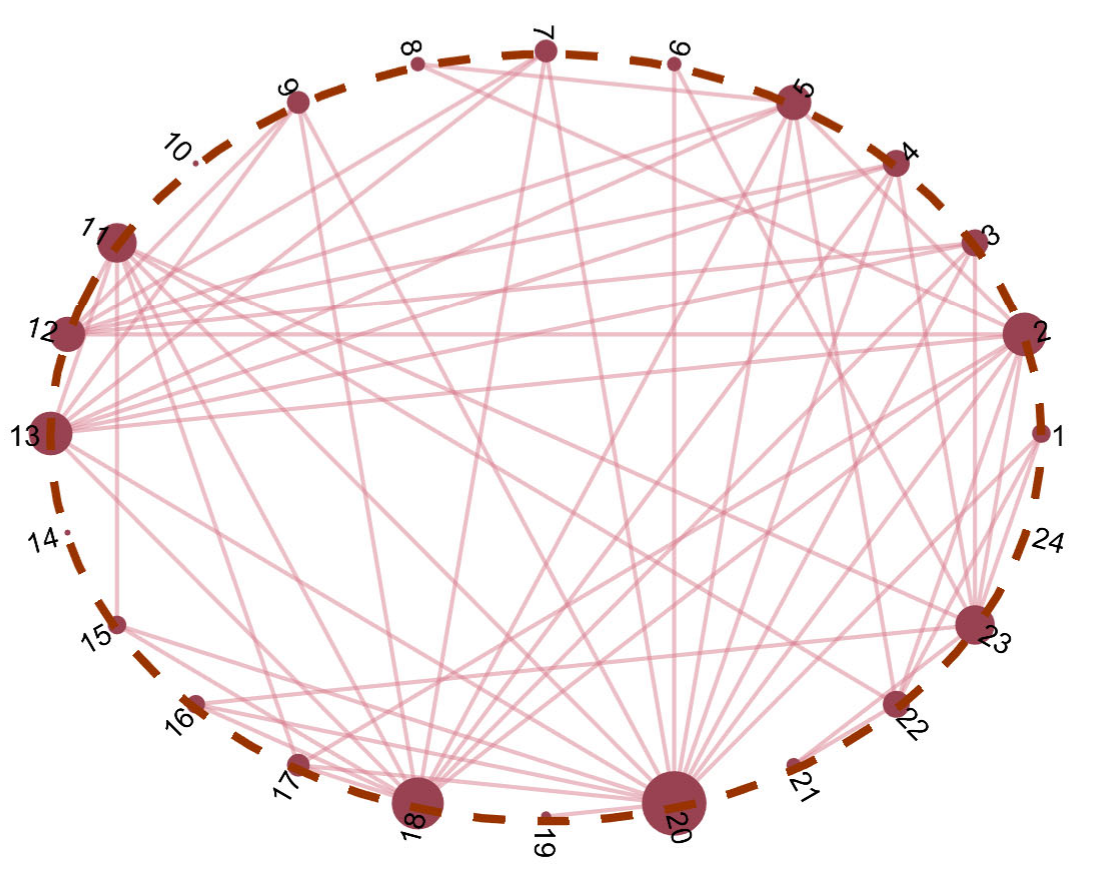}
}
\hfill
\subfloat [Representative graph structure for small-scale events.]{
\includegraphics[width=2.7in]{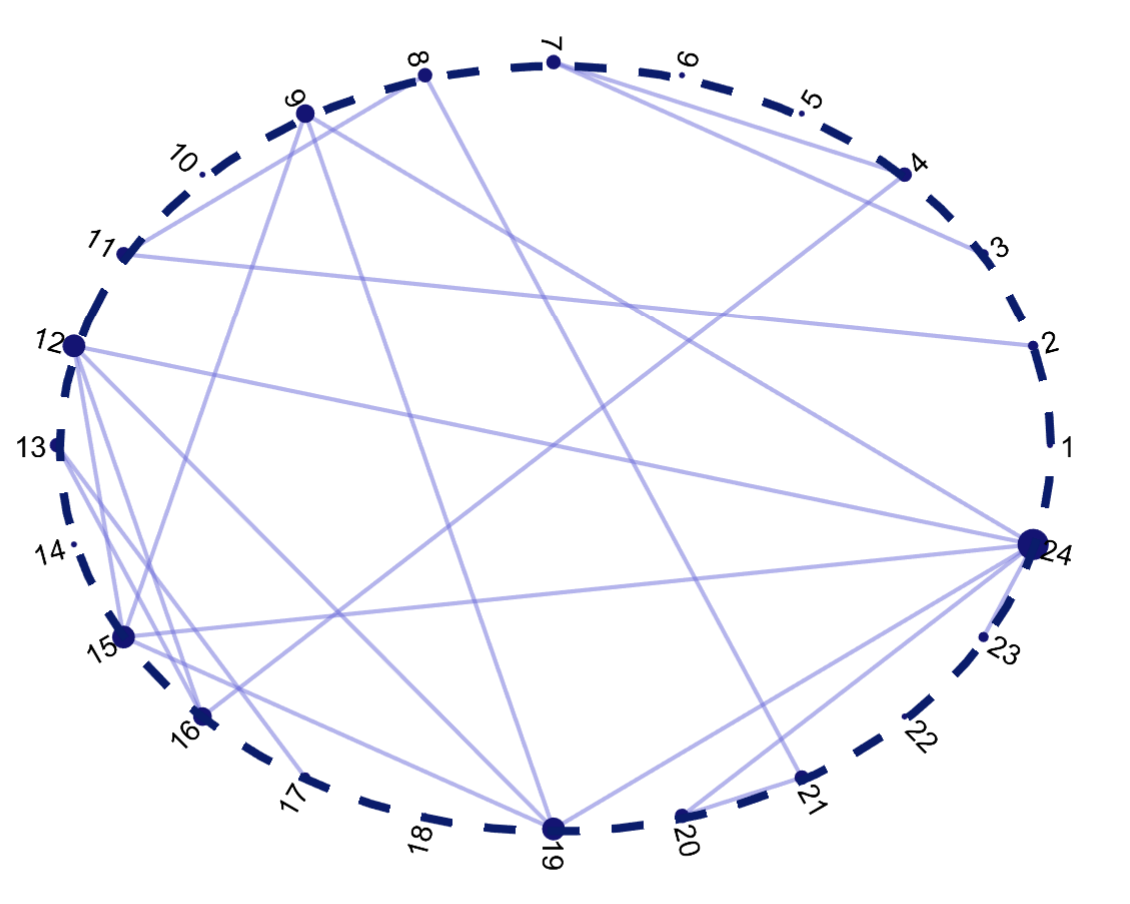}
}
\hfill
\subfloat [Representative graph structure for large-scale events.]{
\includegraphics[width=2.7in]{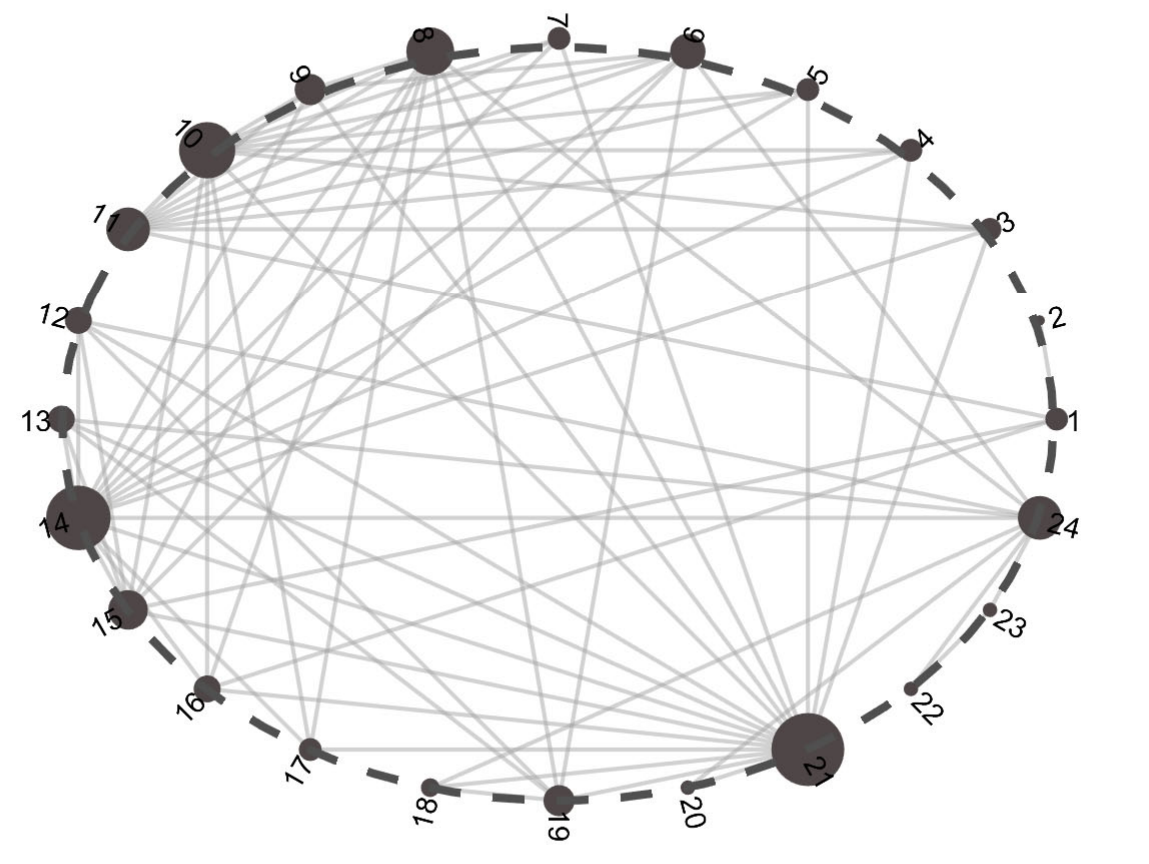}
}
\hfill
\caption{Each representative graph structure (red, green, and blue) corresponds to all data, small-scale events, and large-scale events. The size of a node is proportional to its in-degree.}
\label{fig:rep_figure}
\end{figure}
\subsection{Method Comparison}

We have conducted numerical comparisons with two previous PMU-based event classification models: support vector machine (SVM) \cite{SB2017} and a convolutional neural network (CNN)-based event classification approach \cite{compare}. Also, to further demonstrate the performance of the proposed algorithm, two state-of-the-art classification methods, random forest (RF) and light gradient boosting machine (LGBM) have also been compared with our methods in terms of event classification accuracy using the same dataset \cite{LGBT}. To ensure a fair comparison between the three methods, the performances of the five methods are evaluated based on the same system-level criteria. Specifically, the system-level criteria is calculated as the percentage of times that all PMUs report event type correctly. The hyperparameters of these methods are calibrated by using IBM AutoAI toolkit. As described in Fig. \ref{fig:sample}, the testing accuracy of the proposed method is around 78\%. In contrast, SVM, CNN, RF, and LGBM show the testing accuracy of ${63\%,60\%,61\%,67\%}$, respectively. Hence, based on this real-world PMU dataset, our method outperforms various existing methods. This comparison result also corroborates the premise of this work: investigating interactive relationships among different PMUs is crucial for data-driven event classification tasks.

\subsection{Performance of the Interactive Graph Inference}
Fig. \ref{fig:rep_figure} describes the results of our data-driven interaction inference. In particular, Fig. \ref{fig:rep_figure} shows the representative graph structures with the best performance (i.e., deterministic thresholding, smooth coefficient is 0.5, and data augmentation). Since the graphs are different for each event, we aggregate all the graphs and then select the most frequently appearing (i.e., top 10$\%$) edges as the representation graph structure. Specifically, Fig. \ref{fig:rep_figure}(a) is a representative graph through all training data, which contains the most frequently activated interactions, regardless of the type and size of the events. Fig. \ref{fig:rep_figure}(b) and (c) are representative graphs using small and large-scale events, respectively. It is clear that the connectivity is related to the size of the events. For example, the second graph shows relatively sparse connectivity compared to the first and the third graphs.

In addition to representative graph visualizations, we perform Monte Carlo simulations and measure the dissimilarity of the learned graphs over repeated simulations to evaluate the performance of our method \cite{ST2017}. It should be noted that it is not appropriate to discuss the accuracy of the learned graphs because there is no interactive ground truth. The rationale for the dissimilarity assessment is that a low level of dissimilarity among the learned graphs implies that the learned graphs are valuable and the proposed method is reliable. Here, we utilize a metric, D-measure, to quantify graph dissimilarities between $G_1$ and $G_2$, which is calculated as follows \cite{ST2017}:
\begin{equation}
\begin{split}
\label{eq:dmeasure}
D(G_1,G_2) = &0.45\cdot\sqrt{\frac{J(\mu_{G_1},\mu_{G_2})}{\log2}} + 0.45\cdot|\sqrt{\Phi(G_1)} \\
& -\sqrt{\Phi(G_2)}|+0.05\cdot(\sqrt{\frac{J(\alpha_{G_1},\alpha_{G_2})}{\log2}} \\
& +\sqrt{\frac{J(\alpha_{G^c_1},\alpha_{G^c_2})}{\log2}})
\end{split}
\end{equation}
where, $\alpha_{G_i}$ and $\alpha_{G_i^c}$ are the $\alpha$-centrality values of graph $G_i$ and its complement. $\Phi(G_i)$ is the node dispersion of graph $G_i$, which is defined as follows: 
\begin{equation}
\label{eq:nnd}
\Phi(G_i)=\frac{J_{G_i}(P_1,...,P_N)}{\log(\eta+1)}
\end{equation}
where, $\eta$ is the graph's diameter and $P_i$ is the distance distribution of node $i$ in graph. $J_{G_i}(P_1,...,P_N)$ is calculated from the set of $N$ distance distributions in $G_i$ using the Jensen-Shannon divergence:
\begin{equation}
\label{eq:nnd_2}
J_{G_i}(P_1,...,P_N) = \frac{1}{N}\sum_{i,j}p_i(j)\log(\frac{p_i(j)}{\frac{1}{N}\sum_{i=1}^N p_i(j)})
\end{equation}
Note that $\mu_{G_1} = \frac{1}{N}\sum_{i=1}^N p_i(j)$.

Mathematically, the theoretical lower boundary value of $D_{G_1,G_2}$ is zero; this case happens only when $G_1$ and $G_2$ have the same graph distance distribution, the same graph node dispersion, and the same $\alpha$-centrality vector. In general, a low D-measure indicates that the dissimilarity of the two learned graphs is small. In this work, based on 100 simulations, the average D-measure is relatively low, which is about 0.3. This result shows that the proposed data-driven interaction inference works reliably, and the learned graphs are meaningful. Note that, by analyzing these learned interactive graphs, the proposed method has the potential to be extended in terms of event localization, i.e., finding out the physical location of events in the network. However, since the system topology and historical event locations are not available, we cannot evaluate this work. We leave it to future work once they are available. More comprehensive results will be provided.

\section{Conclusion}\label{conclusion}
In this paper, we have presented a novel solution to accurately and efficiently classify events using all PMU data in the system, without assuming any prior knowledge of the system. Our method establishes on inferring interactive relationships among different PMUs in a data-driven manner. We then embed it into an autoencoder architecture while optimizing graph inference and classification model to significantly improve the performance of the event classifier. Moreover, the proposed framework can automatically capture multi-scale event features with limited parameters by developing a dilated inception model. The scale diversity is enriched by designing paralleled dilated convolutions with various dilation ratios. Numerical experiments using a large-scale real PMU dataset from Western Interconnection show that our data-driven interaction inference works reliably. Also, it is shown that the proposed method can achieve better classification accuracy compared to existing methods.

Future studies will seek to extend the capabilities of the proposed event identification method in two main directions. First, this work has the potential to address the two application challenges mentioned above by investigating unlabeled events and semi-supervised learning techniques. Second, once the system topology and historical event locations are available, we will focus on event localization by exploiting the interaction relationships between different PMUs.

\section*{Acknowledgment and Disclaimer}
Acknowledgment: ``This material is based upon work supported by the Department of Energy under Award Number DE-OE0000910.''

Disclaimer: ``This report was prepared as an account of work sponsored by an agency of the United States Government.  Neither the United States Government nor any agency thereof, nor any of their employees, makes any warranty, express or implied, or assumes any legal liability or responsibility for the accuracy, completeness, or usefulness of any information, apparatus, product, or process disclosed, or represents that its use would not infringe privately owned rights.  Reference herein to any specific commercial product, process, or service by trade name, trademark, manufacturer, or otherwise does not necessarily constitute or imply its endorsement, recommendation, or favoring by the United States Government or any agency thereof.  The views and opinions of authors expressed herein do not necessarily state or reflect those of the United States Government or any agency thereof.''






\ifCLASSOPTIONcaptionsoff
  \newpage
\fi



\bibliographystyle{IEEEtran}
\bibliography{IEEEabrv,./bibtex/bib/IEEEexample}

\begin{IEEEbiography}[{\includegraphics[width=1in,height=1.25in,clip,keepaspectratio]{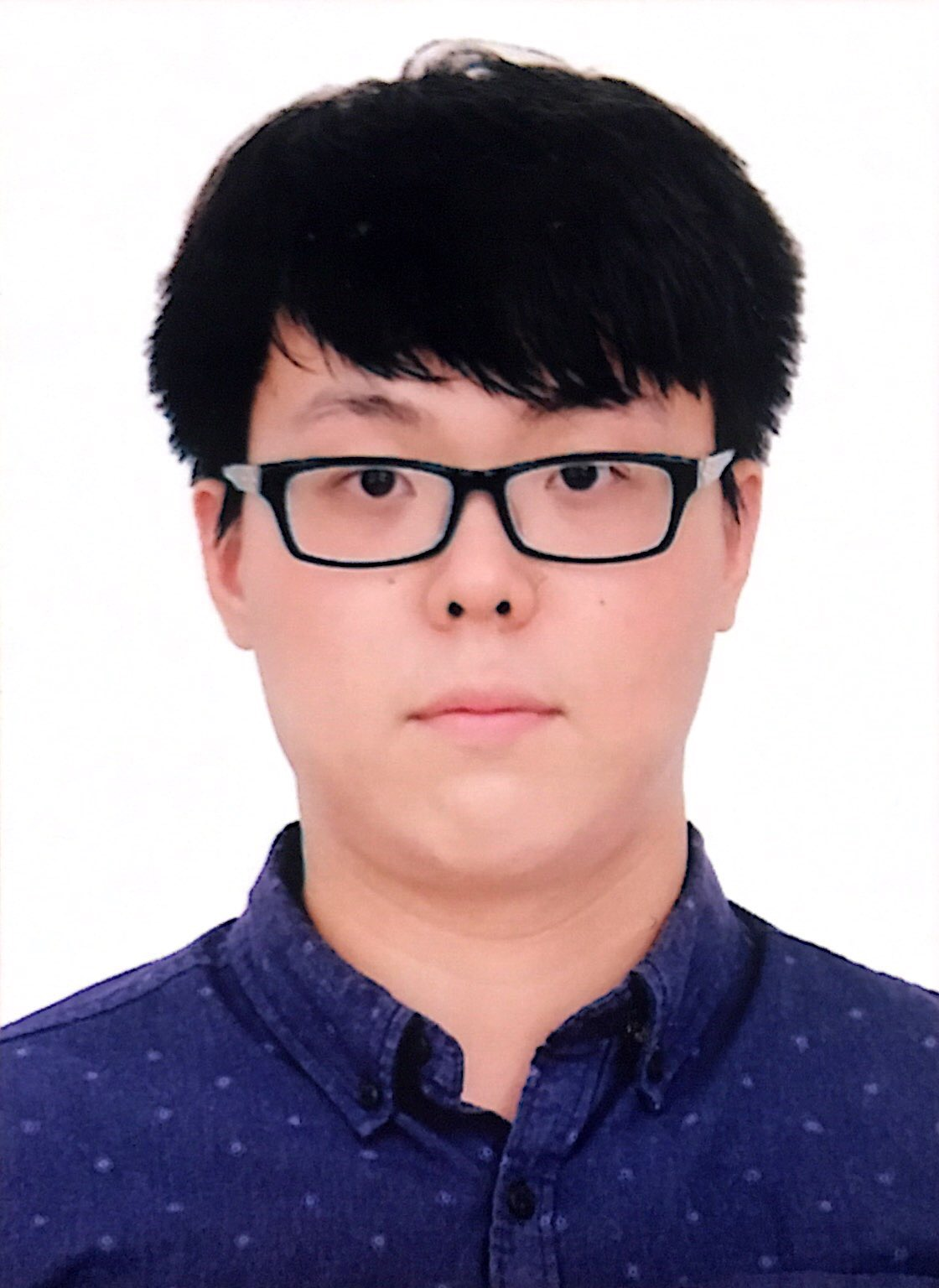}}]{Yuxuan Yuan}(S'18) received the B.S. degree in Electrical \& Computer Engineering from Iowa State University, Ames, IA, in 2017. He is currently pursuing the Ph.D. degree at Iowa State University. His research interests include distribution system state estimation, synthetic networks, data analytics, and machine learning.
\end{IEEEbiography}
\begin{IEEEbiography}[{\includegraphics[width=1in,height=1.25in,clip,keepaspectratio]{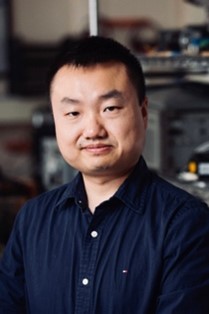}}]{Zhaoyu Wang} (Senior Member, IEEE) received the B.S. and M.S. degrees in electrical engineering from Shanghai Jiaotong University, and the M.S. and Ph.D. degrees in electrical and computer engineering from Georgia Institute of Technology. He is the Northrop Grumman Endowed Associate Professor with Iowa State University. His research interests include optimization and data analytics in power distribution systems and microgrids. He was the recipient of the National Science Foundation CAREER Award, the Society-Level Outstanding Young Engineer Award from IEEE Power and Energy Society (PES), the Northrop Grumman Endowment, College of Engineering’s Early Achievement in Research Award, and the Harpole-Pentair Young Faculty Award Endowment. He is the Principal Investigator for a multitude of projects funded by the National Science Foundation, the Department of Energy, National Laboratories, PSERC, and Iowa Economic Development Authority. He is the Chair of IEEE PES PSOPE Award Subcommittee, the Co-Vice Chair of PES Distribution System Operation and Planning Subcommittee, and the Vice Chair of PES Task Force on Advances in Natural Disaster Mitigation Methods. He is an Associate Editor of IEEE TRANSACTIONS ON POWER SYSTEMS, IEEE TRANSACTIONS ON SMART GRID, IEEE OPEN ACCESS JOURNAL OF POWER AND ENERGY, IEEE POWER ENGINEERING LETTERS, and IET Smart Grid.
\end{IEEEbiography}
\begin{IEEEbiography}[{\includegraphics[width=1in,height=1.25in,clip,keepaspectratio]{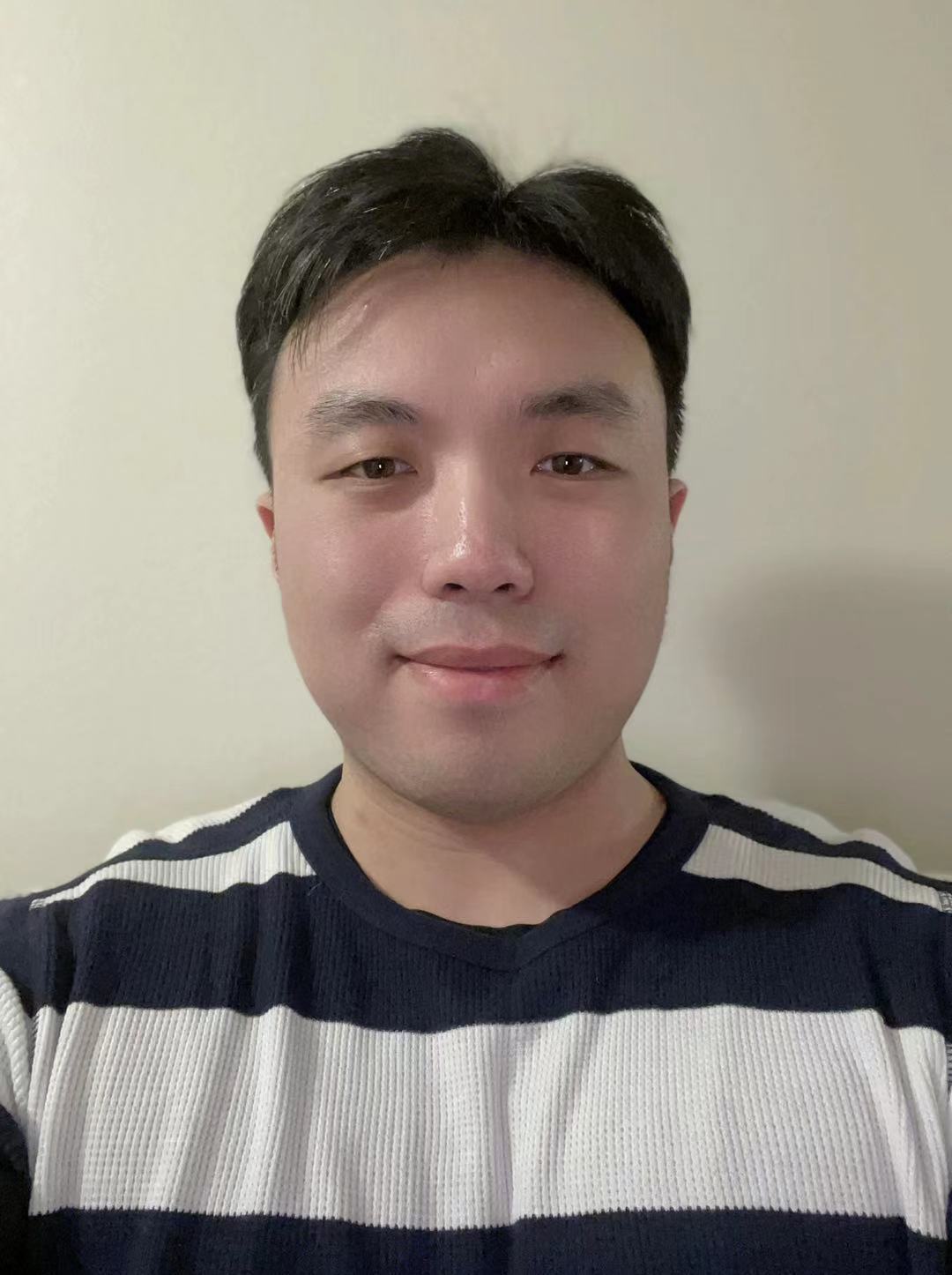}}]{Yanchao Wang} received the Bachelor of Engineering in Optical Information and Technology from Beijing Institute of Technology, Beijing, China in 2014. He is currently pursuing the Ph.D. degree at Iowa State University. His research interests include deep learning in power systems, machine learning and signal processing\end{IEEEbiography}

\end{document}